# Machine learning emulators of dynamic global vegetation models for fast gridded carbon budget predictions

Piyu Ke[1,7*], Xiaofan Gui[2,7], Stephen Sitch[3], Zhu Liu[4], Xun Li[5], Jiang Bian[2], Pierre Friedlingstein[3,6], Philippe Ciais[1,*]

1. Laboratoire des Sciences du Climat et de l'Environnement, University Paris Saclay CEA CNRS, Gif sur Yvette, France
2. Machine learning group, Microsoft Research, Beijing, China
3. Faculty of Environment, Science and Economy, University of Exeter, Exeter, United Kingdom
4. Department of Earth System Science, Ministry of Education Key Laboratory for Earth System Modeling, Institute for Global Change Studies, Tsinghua University, Beijing, China
5. Independent Researcher, Shanghai, China
6. Laboratoire de Météorologie Dynamique, IPSL, CNRS, ENS, Université PSL, Sorbonne Université, École Polytechnique, Paris, France
7. These authors contributed equally: Piyu Ke, Xiaofan Gui.

***Correspondence:** kepiyu@gmail.com (P.K.), philippe.ciais@cea.fr (P.C.)

## Abstract

**The land carbon sink explains most of the variability of the atmospheric $CO_2$ growth rate, but the Global Carbon Budget estimates it with dynamic global vegetation models (DGVMs) run once a year. Emulators could provide it sooner if they reproduce each model outside its training period. Here we show that convolutional neural network emulators of 19 of these models, driven by ERA5 climate and their own past predictions, reproduce the global land sink of these models and its spread over two years not used in training. Global monthly net biome production (NBP) anomalies were reproduced with a median $R^2$ of 0.81. The error of the multi-model mean annual NBP was at most 0.23 GtC $yr^{-1}$. At the grid-cell level, the median $R^2$ of monthly anomalies was 0.30, partly because of errors that accumulate in the emulated NBP history.**

The net carbon flux from the atmosphere to the land is called net biome production (NBP). It is a global carbon sink, which varies from year to year because of climate variability and change. Annual NBP anomalies explain most of the variability of the atmospheric $CO_2$ growth rate[1]. The seasonal cycle of global NBP comes mainly from northern extratropical ecosystems, and its interannual variability mainly from tropical ecosystems in response to El Niño–Southern Oscillation (ENSO) climate modes[2–4]. El Niño events, other droughts, heatwaves and fires are characterized by regional NBP anomalies with weaker carbon uptake, some of which persist for months after the event[5–7].

Dynamic global vegetation models (DGVMs) simulate gridded monthly NBP at resolutions of 0.5° to 1.875° from input 6-hourly gridded climate fields, global annual atmospheric $CO_2$ concentration, and gridded annual land use maps. In the Global Carbon Budget, the land sink ($S_{LAND}$) is the mean NBP simulated by the DGVMs of the TRENDY project[8] in their S2 simulation, in which climate, atmospheric $CO_2$ and nitrogen deposition (for N-enabled models) vary, while the land use map is fixed at its 1700 state[1,8]. Since the 2025 budget, this mean is corrected for historical changes in forest cover, which are not included in S2[9]. The standard deviation among the models is used as the 1σ uncertainty of $S_{LAND}$, about 0.9 GtC $yr^{-1}$ for the year 2024[9]. The TRENDY simulations are run once a year in September, now in the TRENDY v14 edition[9], using the climate forcing of the previous year[10]. Low latency carbon budgets run a small number of DGVMs with operationally updated climate forcing but do not include land use change[11,12].

Machine-learning models trained on climate reanalyses now forecast global weather with an accuracy comparable to that of operational numerical weather prediction[13,14]. Emulators trained on Earth system model simulations predicted the response of temperature and precipitation to emission scenarios[15], and a hybrid model trained on ERA5 simulated the atmospheric climate over several decades[16]. For land carbon models, emulators that take model parameters as inputs have been used to calibrate models against observations and to analyse their sensitivity to parameters[17–19], and an emulator driven by climate has reproduced the forest carbon dynamics of one DGVM[20]. Building and testing emulators of each TRENDY model is the goal of this study.

To be useful, emulators have to meet several requirements. Their error on global annual NBP should be less than 0.4 GtC $yr^{-1}$, the typical error of global NBP inferred from the atmospheric $CO_2$ growth rate[21]. They should reproduce the seasonal cycle and monthly anomalies of NBP, in global totals and in their spatial gridded patterns, and predict NBP anomalies much better than climatology (the mean seasonal cycle of the model), persistence (the last known NBP anomaly of the model held constant) or a simple regression on meteorological anomalies.

Here we train a separate convolutional neural network (CNN) emulator for each of the 19 models of the TRENDY v13 project that provided gridded monthly NBP, remapped to 0.5° by 0.5° where needed, and we emulate their S2 simulations[1,8] (Fig. 1 and Supplementary Table 1). After training and validation during the period 2000–2021, each CNN is run in predictive mode over 2022–2023, driven by ERA5 gridded monthly climate at 0.5° by 0.5°, monthly atmospheric $CO_2$, the vegetation maps of each model, where available, and its own predicted NBP of the preceding 12 months. The TRENDY v13 NBP for 2022–2023 was withheld from training and used only to evaluate the emulators. We also tested emulators trained on 2000–2023 to predict held-out TRENDY v14 gridded NBP for 2024.

# Results

## Emulating each DGVM TRENDY model

Each emulator was trained on the S2 simulation of its target model, one of 18 DGVMs or the data-driven model CARDAMOM[1,8,22,23], and predicts monthly NBP on a 0.5° grid from inputs of the current and preceding 12 months (Fig. 1).

The meteorological inputs of emulators are monthly fields from ERA5[24,25] with eight selected climate variables which are also used in the forcing of the TRENDY models. The ERA5 climate fields are chosen for emulators because they are available about six days after the end of each month, whereas the CRU-JRA long-term climate forcing used by the TRENDY models is only extended once a year[26]. We train the emulators with ERA5 so that the same climate data are used in training and in near-real-time runs. Atmospheric $CO_2$ is from the NOAA Greenhouse Gas Marine Boundary Layer Reference product[27]. The other inputs are the plant functional type (PFT) fractions or gridded vegetation maps of each target model, available for 12 models, the calendar month and location, and the gridded NBP of the preceding 12 months, taken from the target model during training and from the emulator's own predictions when it is run forward. Previous NBP carries legacy information on model state variables, such as carbon pools and leaf area, which pre-condition the prediction (Methods and Supplementary Table 2). The same weights are used at every grid cell, so that one network of about 0.3 million parameters, trained on about 13 million or more grid-cell months per model, represents the response of NBP over the globe, and learned embeddings of location let this response vary geographically (Methods). Each emulator is made of five networks trained on data from 2000 to 2021. The emulator is then run forward from January 2022.

We compared emulated and true NBP in 2022 and 2023. The true NBP of each model in these two years is available in TRENDY v13, but it was not used to train the emulators or as their input when they were run forward. The comparison covers three spatial scales (global, regional and 0.5° grid cell) and three quantities (annual NBP, monthly NBP and monthly NBP anomalies; Supplementary Table 3). Monthly NBP anomalies of each model and of its emulator are deviations from their own 2010–2021 monthly climatology. At each scale, the emulators were compared with climatology, which uses no information after 2021.

## Global NBP

We first ask whether each emulator predicts the global annual NBP of its target model within our requirement of 0.4 GtC $yr^{-1}$. This was the case for 10 of the 19 emulators in 2022 and 6 in 2023, and the median absolute difference between emulated and true global annual NBP was 0.31 GtC $yr^{-1}$ in 2022 and 0.66 GtC $yr^{-1}$ in 2023 (Supplementary Table 4). Climatology was within 0.4 GtC $yr^{-1}$ for 4 models in each year, and its error was larger than that of the emulator for 13 models in 2022 and 15 in 2023.

For the mean of the 19 models, the emulated global annual NBP was 0.08 GtC $yr^{-1}$ lower than the true value in 2022 and 0.23 GtC $yr^{-1}$ higher in 2023. The spread among models is used in the Global Carbon Budget as the 1σ uncertainty of $S_{LAND}$. The standard deviation of annual NBP among the 19 emulators was 1.0 GtC $yr^{-1}$ in 2022 and 1.3 GtC $yr^{-1}$ in 2023, against 1.1 and 1.2 GtC $yr^{-1}$ among the native models, and the $R^2$ between emulated and true annual NBP across the 19 models was 0.68 and 0.61. All 19 models and all 19 emulators simulated lower global NBP in 2023 than in 2022, and the mean decrease was 1.63 GtC $yr^{-1}$ in the models and 1.32 GtC $yr^{-1}$ in the emulators.

We then analyse the ability of emulators to predict the seasonal cycle of global monthly NBP. Global monthly NBP in 2022–2023 shows a strong seasonal cycle, reproduced by emulators with $R^2$ values above 0.92 for all 19 models (median 0.994; Fig. 2,

Supplementary Fig. 1 and Supplementary Table 5). Climatology gave a median $R^2$ of 0.979. The emulators had a higher $R^2$ than climatology for all 19 models. Climatology sets the monthly anomalies to zero, and the emulators reproduce part of them.

Thirdly, as we are interested in temporal variations of NBP, we analyse the ability of emulators to reproduce the more subtle monthly NBP anomalies after removing the mean seasonal cycle. The median $R^2$ of global monthly NBP anomalies over 24 months was 0.81 (Supplementary Table 5). These 24 months include the change from La Niña conditions in 2022 to El Niño conditions in 2023, so this $R^2$ partly reflects a transient decrease in NBP between the two years. We compared the emulators with four baseline predictions of these anomalies: the climatology (zero anomaly), persistence of the December 2021 anomaly, persistence of the mean 2021 anomaly and a regression on ERA5 temperature and precipitation anomalies (Methods). The two persistence baselines set every monthly anomaly of 2022–2023 to the December 2021 anomaly or to the mean 2021 anomaly of the target model. The skill score is one minus the ratio of the mean squared error of the emulator to that of the baseline. A positive skill score indicates that the emulator is closer to the true NBP anomalies than the baseline: a value of one indicates that the emulated anomalies are equal to the true anomalies, and a value of 0.5 indicates that the mean squared error of the emulator is half that of the baseline. The skill score was positive for all 19 models against climatology and both persistence baselines, and for 16 models against the ERA5 regression model. The median skill score was 0.72 against climatology, 0.74–0.78 against the two persistence baselines and 0.47 against the ERA5 regression.

## Regional mean NBP

We then evaluated the emulators' capability to reproduce the spatial patterns of NBP, first for the mean annual NBP over ten large regions from the second phase of the Regional Carbon Cycle Assessment and Processes project (RECCAP2)[28] (Methods). Over the ten regions and the two years, the median absolute difference between emulated and true regional annual NBP was between 0.02 and 0.24 GtC $yr^{-1}$. Over all models and regions, the error of the emulators was smaller than that of climatology in 74% of the model–region cases in 2022 and 54% in 2023. The emulators preserved the differences in regional mean annual NBP between models best in Southeast Asia ($R^2$ of 0.89 across the 19 models) and least in Africa (0.35) (Supplementary Fig. 2).

Regional mean monthly NBP was reproduced with median $R^2$ values from 0.92 in Southeast Asia to 0.995 in North America and Russia, higher than those of climatology in all ten regions. The improvement was largest in Oceania, where the median $R^2$ was 0.92, against 0.71 for climatology. The median ratio of emulated to true amplitude of the mean seasonal cycle of regional NBP was 0.96–1.05 in the ten regions, and the differences in this amplitude between models were preserved ($R^2$ across the 19 models of 0.93–0.998; Supplementary Fig. 2). Regional monthly NBP anomalies were reproduced with median $R^2$ values from 0.32 in Africa to 0.75 in Oceania. These values lie between those for single grid cells (0.30) and for global totals (0.81). Errors at single grid cells may therefore partly cancel in regional and global totals. The skill score relative to climatology had a median above 0.5 in seven of the ten regions and was lowest in Africa (0.30).

## Grid-cell NBP

At the 0.5° grid-cell level, the spatial $R^2$ between the emulated and true maps of mean annual NBP in 2022–2023 had a median of 0.46 (Fig. 3), against 0.21 for the 2010–2021 mean annual NBP map of each model. The spatial $R^2$ values of the emulated maps were higher than those of climatology for 18 of the 19 models, and their root mean square error (RMSE) was lower for all 19. The exception for $R^2$ was CARDAMOM (0.87 against 0.88).

For monthly NBP maps, the $R^2$ between emulated and true NBP was calculated from the time series of each grid cell and averaged over grid cells (Methods). Its median over the 19 models was 0.77, and the median RMSE of grid-level NBP was 7.5 gC $m^{-2}$ $month^{-1}$ (Fig. 3). A baseline NBP prediction model that repeats the 2010–2021 monthly climatology of each model already gave a median $R^2$ of 0.76 for monthly NBP, because monthly NBP at a grid cell is dominated by the seasonal cycle. The mean seasonal cycle of 2022–2023 was reproduced with a median $R^2$ of 0.84, against 0.86 for the 2010–2021 mean seasonal cycle of each model.

Monthly NBP anomalies obtained after removing the seasonal cycle of each grid cell were reproduced less well by the emulators, with median $R^2$ values of 0.30 for monthly and 0.37 for quarterly anomalies (Fig. 3). Climatology, which predicts zero anomaly, had a larger mean squared error than the emulator for all 19 models. The skill score relative to climatology, calculated over all grid cells and months of each model, had a median of 0.23. A skill score of 0.23 means that the mean squared error of the emulator is 23% lower than that of climatology.

The emulators underestimated the amplitude of monthly NBP anomalies. Over the 19 models, the slope of the regression of emulated on true monthly NBP anomalies had a median of 0.34, and the standard deviation of the emulated anomalies was a median of 0.66 times that of the true anomalies (Fig. 4 and Supplementary Figs. 3–5).

Maps of mean annual NBP and of NBP anomalies in January–March (JFM) and July–September (JAS) averaged over 2022 and 2023 show where the emulators differ from the true model values (Fig. 5). For ORCHIDEE and CLM6.0, respectively, the spatial $R^2$ between emulated and true anomaly maps was 0.46 and 0.31 for JFM and 0.15 and 0.17 for JAS, with regression slopes of 0.55 and 0.38 for JFM and 0.29 and 0.29 for JAS (Supplementary Fig. 6). For ORCHIDEE, the largest differences were in tropical South America and Africa. For CLM6.0, they were in northern Eurasia, where the emulator overestimated mean annual NBP and did not reproduce the negative JAS anomalies of the model.

Grid-cell monthly NBP anomalies were reproduced less well in 2023, the second year of prediction, than in 2022, the first year after the training period. The pooled $R^2$ of grid-cell monthly NBP anomalies, calculated over all valid grid cells and months of each year, decreased for all 19 models, from a median of 0.30 in 2022 to 0.22 in 2023 (0.26 for the two years together). When the 19 emulators were run month by month with the NBP of the target model as input instead of their own predictions, this $R^2$ was 0.53 in 2022 and 0.51 in 2023. The lower $R^2$ of the recursive predictions in both years, and especially in 2023, is therefore mainly associated with errors that accumulate in the emulated NBP history.

## Extreme anomalies

The quantile–quantile plots in Fig. 6 compare the distribution of grid-cell monthly NBP anomalies of each target model with that of its emulator. Near zero, the quantiles of the emulators match those of the models, but the departure from the 1:1 line increases towards the 1st and 99th percentiles: the negative anomalies (reduced uptake or net release) of the emulators are less negative, and the positive anomalies less positive, than those of the models. Over the 19 models, the emulators underestimated the 1st percentile of the model anomalies by a median of 37% and the 99th percentile by a median of 34%.

This underestimation is close to that of the standard deviation of the anomalies (0.66 times the true value), so the emulated distribution is narrower as a whole.

SDGVM, ORCHIDEE and EDv3 are examples of small, intermediate and large underestimation of extreme NBP anomalies (Fig. 6): their 1st percentiles were underestimated by 22%, 36% and 43%, and their 99th percentiles by 24%, 31% and 46%. For the other 16 models, the underestimation ranged from 18% (CARDAMOM) to 52% (E3SM) at the 1st percentile and from 8% (CARDAMOM) to 42% (JULES) at the 99th percentile (Supplementary Figs. 7–9).

## Differences among models

The $R^2$ and RMSE between emulated and true NBP differed among the target models (Fig. 3). CARDAMOM had the highest $R^2$ for all five quantities in Fig. 3 (monthly NBP, mean annual NBP, the mean seasonal cycle, and monthly and quarterly anomalies), from 0.41 for monthly anomalies to 0.97 for the mean seasonal cycle, and EDv3 had the lowest $R^2$ for monthly NBP (0.61) and the mean seasonal cycle (0.71). Because the seasonal cycle of NBP was always reproduced better than the anomalies, the $R^2$ for monthly NBP and the mean seasonal cycle was lower for models in which anomalies account for a larger fraction of the monthly variance (Spearman $\rho = -0.73$ and $-0.69$, $P < 0.002$). The spatial $R^2$ of mean annual NBP was higher for models whose grid-cell anomalies were better explained by a linear regression on ERA5 temperature and precipitation anomalies at the same grid cell ($\rho = 0.63$, $P = 0.004$; Methods). This regression describes how closely the anomalies of each model follow local weather and is not one of the baselines. The RMSEs of monthly NBP, the mean seasonal cycle and monthly anomalies increased with the amplitude of the anomalies of the model ($\rho = 0.94$–$0.97$). The $R^2$ values for monthly, quarterly and global anomalies and the underestimation of extreme anomalies were not correlated with any of the model properties examined or with the in-sample $R^2$ of the emulators after adjustment for multiple tests (Methods). The $R^2$ values and the underestimation of extremes did not differ between the 14 models that simulate fire and the 5 that do not ($P \geq 0.18$; Supplementary Table 1), or between the 12 emulators with PFT inputs and the 7 without, after adjustment for multiple tests.

### Prediction for 2024

Differences between the emulators and TRENDY v14 in the held-out year 2024 combine emulator errors with changes between model versions (Methods). For 2024, the emulators trained on 2000–2023 predicted a mean global NBP anomaly of −0.46 GtC $yr^{-1}$ over 17 models, compared with −0.68 GtC $yr^{-1}$ in TRENDY v14 (Supplementary Table 6). The sign of the anomaly was correct for 13 of the 17 models, and 16 of the 17 TRENDY v14 anomalies were negative. For the 17-model mean, the skill score for global monthly anomalies was 0.66 relative to climatology, 0.59 relative to the ERA5 regression and 0.21 relative to persistence of the mean 2023 anomaly of TRENDY v13. For the mean of the ten models whose anomalies were highly correlated between versions (Methods), the skill scores were 0.64, 0.72 and 0.55. For $R^2$ thresholds from 0.85 to 0.95 (12 to 4 models), the skill score relative to persistence ranged from 0.25 to 0.55 (Supplementary Table 6). The $R^2$ of global monthly anomalies was 0.31 for both means. The skill scores relative to climatology are nevertheless positive because the emulators reproduce part of the negative mean anomaly of 2024, which climatology sets to zero. For single models, the median $R^2$ of global monthly anomalies was 0.34, and the skill score relative to climatology was negative for 5 of the 17 models (Supplementary Table 6).

## Discussion

Part of the error in grid-cell NBP anomalies can come from the different climate forcing used by emulators. The emulators take ERA5 inputs but are trained on NBP simulated with the CRU-JRA forcing[10]. Differences between the two forcings act as errors in the inputs. They lower the accuracy of the emulators and can reduce the amplitude of the response learned in training. Over 2010–2022, the median correlation between monthly anomalies of ERA5 and CRU-JRA version 2.4 at single grid cells was 0.89 for temperature and 0.60 for precipitation. For precipitation, it was lowest in Africa (0.28), the region where the emulators reproduced NBP anomalies least well. In 36 grid cells of western Venezuela, CRU-JRA has a precipitation deficit in 2021 that is absent from ERA5 (Supplementary Fig. 10), and these grid cells were excluded from the evaluation (Methods). Emulators trained and run with CRU-JRA inputs would show how much of the error comes from the difference between the two forcings.

The emulators are trained on the TRENDY simulations, but once trained they can be run with ERA5 for all 19 models without new simulations by the modelling groups. Most of the computing cost of an emulator is in training. The five networks of one emulator took about 100–680 GPU hours to train, and retraining all 19 emulators at each TRENDY release would take 3,000–5,400 GPU hours. A trained emulator predicts 12 months in about 2 GPU hours. For comparison, one simulated year of ORCHIDEE at 0.5° costs about 250 CPU core hours, and an ORCHIDEE simulation from 1700 to 2023 about 80,000 core hours without spin-up. GPU hours and CPU core hours are not directly comparable.

At the grid-cell level, the $R^2$ of monthly NBP was lower for models in which anomalies account for a larger fraction of the monthly variance. The median of this fraction over grid cells was 0.37 for EDv3, one of the models that simulate vegetation demography[29], against 0.02 for CARDAMOM. In EDv3, anomalies can arise from mortality, disturbance and fire, which respond to thresholds and to conditions in previous years[7], whereas the meteorological inputs of the emulators cover only the current and preceding 12 months. Across the 19 models, however, the $R^2$ of NBP anomalies and the underestimation of extremes were not related to any of the model properties examined, including the lag-1 autocorrelation of anomalies (Methods).

Grid-cell NBP anomalies were reproduced less well in the second year of prediction (2023) than in the first year (2022). Training on recursive predictions over several months, as done for machine-learning weather models[13], could reduce the errors that accumulate in the emulated NBP history. Retraining the emulators at each TRENDY release shortens the recursion and is also needed because the models and their forcing change between releases (Supplementary Table 6).

We emulated only the S2 simulation, in which land use is fixed at its 1700 state[1,8]. S2 NBP excludes emissions from land-use change ($E_{LUC}$). It also includes the sink of forests that were cleared after 1700 but are kept in S2. This sink is called the loss of additional sink capacity[30]. S2 NBP therefore cannot be compared directly with the net land flux estimated by atmospheric inversions. Since the 2025 budget, the global $S_{LAND}$ of the DGVMs is reduced by 19% to account for the sink of these cleared forests[9], and the same correction can be applied to the emulated $S_{LAND}$. NBP with land-use change could be emulated by training on the S3 simulation, in which land use also varies. Such an emulator would need land-use states and transitions as inputs, because new deforestation and harvest are not carried by the NBP history, and their effects on NBP last for years. Land-use transitions after the last year of the TRENDY land-use forcing[31] could be taken from satellite forest-loss alerts[32]. The emulated $S_{LAND}$ could also be combined with the bookkeeping $E_{LUC}$ that the Global Carbon Budget projects for the current year[1].

Fire emissions depend non-linearly on fuel accumulated over years and on fire weather lasting a few days. Monthly mean inputs do not resolve such short events. Only five models do not simulate fire, so the comparison between models with and without fire in the Results has little power. Monthly counts of days with high fire danger from ERA5[33] could be added as inputs. Fire emissions could also be emulated separately, with more weight on rare large fires in the loss function used for training. Emulating gross primary production and ecosystem respiration separately as well, with NBP as the balance of these and the other carbon fluxes of each model, would show which flux drives an NBP anomaly.

A memory longer than 12 months, for example in recurrent networks[34], could capture multi-year legacies of droughts and disturbances. Carbon pools and leaf area could also be predicted with NBP and used as inputs.

The smaller amplitude of the emulated anomalies is consistent with training on a squared error, which drives predictions towards the conditional mean. The conditional mean varies less than the target when the inputs do not fully determine it[35]. Oversampling rare months or probabilistic training[36] may preserve more of the extremes.

Annual biases contribute little to the mean squared error of monthly grid-cell NBP that is minimized in training. A bias of 1 GtC $yr^{-1}$ spread over all grid cells is below 1 gC $m^{-2}$ $month^{-1}$, against a median RMSE of 7.5 gC $m^{-2}$ $month^{-1}$. The recursive training discussed above would also allow a loss term on annual totals to reduce these biases. Hybrid models, in which process equations constrain part of the learned response[37], may be more robust outside the training period.

# Methods

We trained a separate emulator for each of 19 TRENDY models. Each emulator predicts the monthly NBP of its target model on a 0.5° grid from meteorological and $CO_2$ predictors and the NBP of the preceding 12 months, so that it can be run forward month by month once the predictors are available. The emulators were trained on 2000–2021 and tested on 2022–2023. A second set, trained on 2000–2023, was tested on 2024 (see Out-of-sample tests).

## Target data

We used monthly net biome production (NBP) fields from simulation 2 (S2) of TRENDY v13, the model ensemble of the Global Carbon Budget 2024. S2 is forced by time-varying atmospheric $CO_2$, climate and, in models with a nitrogen cycle, nitrogen deposition, with land use fixed at its 1700 state, and is the simulation used in the Global Carbon Budget to estimate the land sink[1,8]. The targets were the 19 models that provided monthly NBP for 2000–2023 (2003–2023 for CARDAMOM; Supplementary Table 1): 18 dynamic global vegetation models (DGVMs) and the data-driven model CARDAMOM[22,23].

NBP fields not provided on a 0.5° × 0.5° grid were remapped to this grid by conservative remapping with Climate Data Operators (CDO), which preserves area-integrated fluxes[38]. NBP was provided in kgC $m^{-2}$ $s^{-1}$. Positive values denote net land carbon uptake. For grid-cell statistics, monthly NBP was converted to gC $m^{-2}$ $month^{-1}$ using the number of days in each month, and annual values are sums over calendar years. For global totals, grid-cell fluxes were multiplied by grid-cell area and expressed in GtC $month^{-1}$ or GtC $yr^{-1}$.

## Predictors

For each target month, the inputs at each 0.5° grid cell were eight meteorological variables and atmospheric $CO_2$ in that month and the preceding 12 months, the NBP of the target model in the preceding 12 months, the plant functional type (PFT) fractions of the target model where available, and the calendar month, latitude and longitude. The meteorological variables were ERA5 monthly means of surface pressure, downward shortwave and longwave radiation, 2 m air temperature, total precipitation, specific humidity at 1000 hPa, and the zonal and meridional components of 10 m wind (Supplementary Table 2)[24,25,39]. All eight variables are also provided by CRU-JRA, the climate forcing of the TRENDY simulations, but CRU-JRA is updated only once a year. Version 2.5, used in TRENDY v13, covers 1901–2023[10]. ERA5 is updated daily about five days behind real time, and its monthly means are available around the 6th of the following month[24]. Because the emulators are trained on simulations forced with CRU-JRA, they estimate the NBP that each model would simulate with CRU-JRA forcing, using ERA5 as a substitute for CRU-JRA. ERA5 was compared with CRU-JRA version 2.4, the forcing of TRENDY v12, through the correlation of their monthly anomalies of 2 m air temperature and precipitation at each grid cell over 2010–2022. The ERA5 fields were remapped from 0.25° to a 0.5° grid by bilinear interpolation (cdo remapbil)[38].

Atmospheric $CO_2$ was taken from the NOAA Greenhouse Gas Marine Boundary Layer Reference product, averaged to monthly values and interpolated linearly from its 41 latitude bands to 1° latitude[27]. The $CO_2$ predictor therefore varied with latitude and month but not with longitude. Over the 13 months of each input window, this $CO_2$ field also contains the observed $CO_2$ growth rate, which depends on the real land sink. The TRENDY models were forced with global annual $CO_2$, so the emulators could take part of the NBP anomalies from an observed signal that the models did not have. The calendar month, latitude and longitude were included as learned embeddings to represent seasonality and spatial position. The year was not an input. The PFT fractions of each model describe its vegetation distribution, on which the response of NBP to weather depends, and were used for the 12 models that provided them, with one input channel per PFT. In 2022 and 2023, the PFT fractions of the same year were used, as in training. They differ from those of 2021 only for JULES, LPX-Bern and SDGVM. The other seven models had no PFT input, and no external land-cover map was used in their place. The NBP of the preceding 12 months gives the recent flux of the target model at each grid cell, which depends on model state variables that are not inputs, such as carbon pools and leaf area.

## Emulator architecture

Each emulator consists of five convolutional neural networks with the same architecture (see Training). Each network predicts NBP in the target month on a patch of 18 × 18 grid cells (9° × 9°) from the inputs of that month and the preceding 12 months (Fig. 1). For each month, the numerical inputs of each grid cell (meteorological variables, $CO_2$, NBP and, where used, PFT fractions) are mapped by a linear layer to 64 features, to which learned embeddings of the calendar month, latitude and longitude are added. The NBP input of the target month is masked. The features of the 13 months are concatenated (832 channels) and reduced to 64 channels by a 1 × 1 convolution. Five residual blocks[40] follow, each consisting of a 3 × 3 convolution with 64 channels, group normalization with eight groups[41], a GELU activation[42] and dropout with a rate of 0.25[43], with a skip connection around the block. A two-layer perceptron with 64 hidden units and GELU activation, applied to each grid cell, gives NBP in the target month.

The 13-month input window allows the emulator to represent lagged responses of NBP to weather within a year, for example through soil moisture or the effects of drought on vegetation in subsequent months. Longer input windows increased the computing cost and memory use without improving the results. The convolutions apply the same weights at all locations, so that one set of weights describes the response to the inputs in all grid cells, while the location embeddings allow this response to vary geographically. With five 3 × 3 convolutions, the convolutional receptive field of each grid cell is 11 × 11 grid cells (5.5° × 5.5°). Each network has 313,473 trained parameters, plus 64 for each PFT channel (315,393 for LPX-Bern, which has 30 PFTs).

## Training

Inputs and targets, except the PFT fractions, were standardized with the RobustScaler of scikit-learn[44], fitted to the 2000–2021 data of each target model, and missing input values were set to zero after standardization. Samples were drawn from 18 × 18 patches placed with a stride of nine grid cells, so that adjacent patches overlap by half, and from 13-month windows shifted by one month. All patch positions were used, including those without land. For 2000–2021, this gave 776,412 samples per target model (3,081 patch positions × 252 windows). These samples contained 12.9 million or more grid-cell months with target values, each counted once, more than 40 times the number of trained parameters.

The networks were trained by minimizing $L = L_{sup} + 0.2\, L_{con}$, with both terms computed in every optimization step. In each step, two masked copies of the standardized numerical inputs were generated by setting each value to zero with a probability of 10% (weak masking) or 30% (strong masking). $L_{sup}$ is the mean squared error between the prediction from the weakly masked inputs and the target-model NBP, calculated over grid cells with target values. $L_{con}$ is the mean squared error between the predictions from the strongly and weakly masked inputs, calculated over grid cells without target values, which include ocean grid cells and grid cells outside the land mask of the target model. This consistency term follows the weak and strong augmentation scheme of FixMatch[45] and was included to stabilize the predictions along coastlines and mask edges, where patches contain few target values.

The samples were divided at random into five folds (KFold of scikit-learn, with shuffling), and five networks were trained, each with a different fold held out for validation. Because patches and windows overlap, the grid-cell months of a validation fold also occur in the training samples of the same network. The validation loss was used only to stop training and to select the network, and the emulators were evaluated on years withheld from training (next section). Each network was trained with AdamW[46] at a constant learning rate of $3.2 \times 10^{-5}$ and a weight decay of $10^{-3}$, a batch size of 512 samples (128 on each of four GPUs) and gradient clipping at 5. Training stopped when the validation loss had not decreased for ten epochs, with a maximum of 1,000 epochs, and the network with the lowest validation loss was kept. The networks were trained for about 60–510 epochs. The output of an emulator is the mean of its five networks. The predictions of overlapping patches were averaged at each grid cell. The emulators were implemented in PyTorch[47].

## Out-of-sample tests

Target-model NBP for 2022–2023 was withheld from training and used to test the emulators. From January 2022, each emulator was run recursively: the NBP predicted for each month was used as NBP input for the following months, so that no target-model NBP after December 2021 entered the predictions. For 2010–2021, emulator output was calculated month by month with the target-model NBP as input. This output is in-sample, because each grid-cell month of this period was in the training samples of at least four of the five networks. For all 19 emulators, output for 2022–2023 was also calculated month by month with the target-model NBP as input.

A second set of emulators was trained in the same way on TRENDY v13 data for 2000–2023 and run recursively from January 2024, with the PFT fractions of 2023. Their predictions for 2024 were compared with S2 of TRENDY v14, the model ensemble of the Global Carbon Budget 2025[9]. OCN and ISBA-CTRIP are not in TRENDY v14. LPJwsl is named LPJ-EOSIM in TRENDY v14[9]. The 2024 test therefore includes 17 models. Because the differences between the emulators and TRENDY v14 combine emulator errors with changes between model versions, we also report means over the ten models whose global monthly NBP anomalies over 2010–2021 were highly correlated between versions ($R^2$ between TRENDY v13 and v14 above 0.88; Supplementary Table 6).

## Evaluation

For each target model, the evaluation used the grid cells with complete NBP for 2010–2023 and a temporal standard deviation above $10^{-10}$ kgC m$^{-2}$ s$^{-1}$. No common land mask was applied, so the number of grid cells differs among models. We also excluded 36 grid cells in western Venezuela (6.75–9.75° N, 73.75–70.25° W), where the CRU-based forcing shows a decrease in precipitation in 2021 that is absent from ERA5 (Supplementary Fig. 10). In these grid cells, annual precipitation in 2021 was 0.32–0.33 times the 2010–2020 mean in CRU-JRA v2.3 and v2.4 (TRENDY v11 and v12)[26,48] and 0.40 times in CRU TS 4.08[49], on which CRU-JRA v2.5 (TRENDY v13) is based, compared with 0.92 times in ERA5. In TRENDY v11–v13, LPX-Bern and LPJwsl simulated fire emissions in 2021 in these grid cells more than 20 times their 2010–2020 mean. The 36 grid cells are the connected region in which 2021 precipitation in CRU-JRA v2.4 was below half of the 2010–2020 mean. They were used in training but excluded from the evaluation.

$R^2$ is the squared Pearson correlation coefficient. Anomalies are deviations from the 2010–2021 monthly climatology of each dataset (model or emulator). Quarterly anomalies are means of the monthly anomalies over January–March, April–June, July–September and October–December. The mean seasonal cycle is the 2022–2023 mean for each calendar month, and mean annual NBP is the 2022–2023 mean. The amplitude of the mean seasonal cycle is its maximum minus its minimum. Climatology, the baseline used at all scales, repeats the 2010–2021 values of the target model: its mean annual NBP for annual totals, its mean for each calendar month for monthly NBP, its mean annual NBP map for the spatial comparison of mean annual NBP, and zero for monthly anomalies.

Grid-cell statistics were calculated from the time series at each valid grid cell and summarized by the mean $R^2$ and the median RMSE over grid cells (Fig. 3). Spatial statistics were calculated across valid grid cells between maps of the emulator and the model (mean annual NBP in Fig. 3, Fig. 5 and Supplementary Fig. 6). Pooled statistics were calculated from all valid grid cells and months together, over 2022–2023 (Fig. 4 and Supplementary Figs. 3–5) or for each year separately. Global statistics were calculated from global totals over valid grid cells (Fig. 2, Supplementary Fig. 1 and Supplementary Tables 4–6). Regional statistics were calculated from totals over ten regions based on the RECCAP2 regions[28] and defined from a 0.5° country mask: North America (United

States, Canada, Greenland, Iceland, Svalbard and the Faroe Islands), Europe, West Asia, Russia, East Asia, South Asia, Southeast Asia, Latin America (South America, Mexico, Central America and the Caribbean), Oceania (Australia and New Zealand) and Africa. Regional totals were summed over the valid grid cells of each model, so the area of a region differs among models.

Quantile–quantile plots compare the quantiles of the emulator and model anomalies, pooled over valid grid cells and the 24 months of 2022–2023 (Fig. 6 and Supplementary Figs. 7–9). The underestimation of extreme anomalies was quantified at the 1st and 99th percentiles as one minus the ratio of the emulator percentile to the model percentile.

The skill score of an emulator relative to a baseline prediction is $SS = 1 - MSE_{em}/MSE_{base}$, where $MSE_{em}$ and $MSE_{base}$ are the mean squared errors of the emulator and of the baseline with respect to the monthly anomalies of the target model. SS is positive when the emulator is closer to the target model than the baseline. SS was calculated from the 24 global or regional monthly anomalies of 2022–2023, and at the grid-cell level from all valid grid cells and months of 2022–2023 together, without area weighting. Regional and grid-cell anomalies were compared with climatology only. For global anomalies in 2022–2023 (24 months; Supplementary Table 5), four baselines were used: climatology (zero anomaly), persistence of the December 2021 anomaly of the target model, persistence of its mean anomaly in 2021 and a linear regression on ERA5. Climatology tests whether the emulators add information to the mean seasonal cycle of the target model. Persistence tests whether they do better than assuming that the last known anomaly continues. The regression tests whether they do better than a linear model of the response of global NBP to temperature and precipitation. The regression predicted the global monthly anomaly of the target model from eight predictors: anomalies of 2 m air temperature and of precipitation averaged over tropical ($|\text{latitude}| < 23.5°$) and extratropical land, each in the current month and averaged over the three preceding months. Land was defined by the RECCAP2 regions[28]. The regression was fitted by ordinary least squares to 2010–2021 and applied to ERA5 data for 2022–2023. For 2024 (12 months; Supplementary Table 6), the target was TRENDY v14, persistence was the mean 2023 anomaly in TRENDY v13, and the regression was fitted to TRENDY v13 for 2010–2023.

Multi-model means were calculated by averaging the global monthly series of the models, and separately those of their emulators, before calculating the statistics. The annual bias is the global annual NBP of the emulator minus that of the model. For 2022 and 2023, the bias was compared with the 2010–2021 mean bias ± 2 standard deviations (Supplementary Table 4). Because the 2010–2021 emulator output is in-sample, this range reflects the fit to the training data. The 2024 annual anomaly is the 2024 global total minus the 2010–2021 mean of the same dataset.

Differences in $R^2$, RMSE and the underestimation of extremes among the 19 emulators were compared with properties of the target models, computed from model output and ERA5, and with the in-sample $R^2$ of the emulators, using Spearman rank correlations across models. The properties were the standard deviation of monthly anomalies in 2010–2021 and 2022–2023 (pooled over grid cells and months, and the median over grid cells of the standard deviation at each grid cell), and the 1st and 99th percentiles of the pooled anomalies in 2022–2023 and their difference. For 2010–2021, the properties also included the medians over grid cells of the fraction of monthly variance due to anomalies, of the lag-1 autocorrelation of monthly anomalies and of the $R^2$ of an ordinary least-squares regression of monthly anomalies on ERA5 2 m air temperature and precipitation anomalies at the same grid cell, in the current month and averaged over the three preceding months. We also used the number of valid grid cells, the grid spacing of the model, the number of PFT channels and, for the models that provide fire emissions, the interannual standard deviation and coefficient of variation of global fire emissions. The in-sample $R^2$ was the pooled $R^2$ of grid-cell monthly anomalies in 2021. The $P$ values given in Results are unadjusted. The correlations reported there have Benjamini–Hochberg adjusted $P$ values below 0.05 over all correlations tested. Models with and without PFT inputs, and with and without fire, were compared with two-sided Mann–Whitney U tests, with Benjamini–Hochberg adjustment over the 30 tests.

## Computational cost

Each emulator was trained on four NVIDIA V100 GPUs. Training the five networks of an emulator took about 26–59 h (about 100–240 GPU hours), and 160–170 h (640–680 GPU hours) for LPX-Bern and ISBA-CTRIP, which have the most PFT channels (30 and 19), or 3,000–5,400 GPU hours for the 19 emulators. A 12-month recursive prediction took about 30 min per emulator on four V100 GPUs, including data loading (about 2 GPU hours), or about 38 GPU hours for the 19 emulators, which can run in parallel. For comparison, one simulated year of a global ORCHIDEE simulation at 0.25° resolution with 6-hourly forcing takes about 1,000 CPU core hours on 44 cores, excluding spin-up.

The cost per land grid cell and simulated year was similar in a simulation at 2° (3–4 × $10^{-3}$ core hours), so one simulated year at 0.5°, with four times fewer land grid cells than at 0.25°, costs about 250 core hours, and a simulation from 1700 to 2023 about 80,000 core hours without spin-up. Once trained, an emulator thus predicts one year in about 30 min on four GPUs, whereas ORCHIDEE at 0.5° would take about 6 h on 44 CPU cores.

### Use of AI tools

The AI language models Claude (Anthropic) and ChatGPT (OpenAI) were used to help edit the text and to help improve the data-processing code. The authors designed the study, checked all analyses and text, and take full responsibility for the manuscript.

## Data availability

The TRENDY v13 and v14 S2 simulations, including fire emissions and land-cover fractions, are available from the Global Carbon Budget data portal (https://mdosullivan.github.io/GCB/) under the TRENDY data policy. ERA5 monthly data are available from the Copernicus Climate Data Store (https://doi.org/10.24381/cds.f17050d7 and https://doi.org/10.24381/cds.6860a573). The NOAA Greenhouse Gas Marine Boundary Layer Reference $CO_2$ product is available at https://doi.org/10.15138/DVNP-F961. CRU-JRA and CRU TS data are available from the Centre for Environmental Data Analysis (https://catalogue.ceda.ac.uk). The RECCAP2 region mask is available from the RECCAP2 project[28]. The emulated monthly NBP for 2022–2024, the evaluation mask and the data underlying the figures and tables are available at Zenodo (https://doi.org/10.5281/zenodo.23677085).

## Code availability

The code used to train and run the emulators and to produce the figures and tables is available at https://github.com/agiamason/NRT_LAND and archived at Zenodo (https://doi.org/10.5281/zenodo.23681671).

## References

1. Friedlingstein, P. *et al.* Global Carbon Budget 2024. *Earth Syst. Sci. Data* **17**, 965–1039 (2025).
2. Graven, H. D. *et al.* Enhanced Seasonal Exchange of $CO_2$ by Northern Ecosystems Since 1960. *Science* **341**, 1085–1089 (2013).

3. Poulter, B. *et al.* Contribution of semi-arid ecosystems to interannual variability of the global carbon cycle. *Nature* **509**, 600–603 (2014).
4. Ahlström, A. *et al.* The dominant role of semi-arid ecosystems in the trend and variability of the land $CO_2$ sink. *Science* **348**, 895–899 (2015).
5. Cox, P. M. *et al.* Sensitivity of tropical carbon to climate change constrained by carbon dioxide variability. *Nature* **494**, 341–344 (2013).
6. Liu, J. *et al.* Contrasting carbon cycle responses of the tropical continents to the 2015–2016 El Niño. *Science* **358**, eaam5690 (2017).
7. Frank, D. *et al.* Effects of climate extremes on the terrestrial carbon cycle: concepts, processes and potential future impacts. *Glob. Change Biol.* **21**, 2861–2880 (2015).
8. Sitch, S. *et al.* Trends and Drivers of Terrestrial Sources and Sinks of Carbon Dioxide: An Overview of the TRENDY Project. *Glob. Biogeochem. Cycles* **38**, e2024GB008102 (2024).
9. Friedlingstein, P. *et al.* Global Carbon Budget 2025. *Earth Syst. Sci. Data* **18**, 3211–3288 (2026).
10. University of East Anglia Climatic Research Unit & Harris, I. C. CRU JRA v2.5: a forcings dataset of gridded land surface blend of Climatic Research Unit (CRU) and Japanese reanalysis (JRA) data; Jan. 1901-Dec. 2023. NERC EDS Centre for Environmental Data Analysis (2024).
11. Ke, P. *et al.* Low latency carbon budget analysis reveals a large decline of the land carbon sink in 2023. *Natl. Sci. Rev.* **11**, nwae367 (2024).
12. Ciais, P. *et al.* Low latency global carbon budget indicates reduced land carbon sink in the year 2024. *Natl. Sci. Rev.* **13**, nwaf594 (2026).
13. Lam, R. *et al.* Learning skillful medium-range global weather forecasting. *Science* **382**, 1416–1421 (2023).
14. Bi, K. *et al.* Accurate medium-range global weather forecasting with 3D neural networks. *Nature* **619**, 533–538 (2023).
15. Watson‐Parris, D. *et al.* ClimateBench v1.0: A Benchmark for Data‐Driven Climate Projections. *J. Adv. Model. Earth Syst.* **14**, e2021MS002954 (2022).
16. Kochkov, D. *et al.* Neural general circulation models for weather and climate. *Nature* **632**, 1060–1066 (2024).

17. Dagon, K., Sanderson, B. M., Fisher, R. A. & Lawrence, D. M. A machine learning approach to emulation and biophysical parameter estimation with the Community Land Model, version 5. *Adv. Stat. Climatol. Meteorol. Oceanogr.* **6**, 223–244 (2020).
18. Baker, E., Harper, A. B., Williamson, D. & Challenor, P. Emulation of high-resolution land surface models using sparse Gaussian processes with application to JULES. *Geosci. Model Dev.* **15**, 1913–1929 (2022).
19. Fer, I. *et al.* Linking big models to big data: efficient ecosystem model calibration through Bayesian model emulation. *Biogeosciences* **15**, 5801–5830 (2018).
20. Natel, C. *et al.* Emulating grid-based forest carbon dynamics using machine learning: an LPJ-GUESS v4.1.1 application. *Geosci. Model Dev.* **18**, 4317–4333 (2025).
21. Friedlingstein, P. *et al.* Emerging climate impact on carbon sinks in a consolidated carbon budget. *Nature* **649**, 98–103 (2026).
22. Bloom, A. A., Exbrayat, J.-F., Van Der Velde, I. R., Feng, L. & Williams, M. The decadal state of the terrestrial carbon cycle: Global retrievals of terrestrial carbon allocation, pools, and residence times. *Proc. Natl. Acad. Sci.* **113**, 1285–1290 (2016).
23. Smallman, T. L. *et al.* Parameter uncertainty dominates C-cycle forecast errors over most of Brazil for the 21st century. *Earth Syst. Dyn.* **12**, 1191–1237 (2021).
24. Hersbach, H. *et al.* ERA5 monthly averaged data on single levels from 1940 to present. Copernicus Climate Change Service (C3S) Climate Data Store (CDS) https://doi.org/10.24381/CDS.F17050D7 (2023).
25. Hersbach, H. *et al.* The ERA5 global reanalysis. *Q. J. R. Meteorol. Soc.* **146**, 1999–2049 (2020).
26. University of East Anglia Climatic Research Unit & Harris, I. C. CRU JRA v2.4: a forcings dataset of gridded land surface blend of Climatic Research Unit (CRU) and Japanese reanalysis (JRA) data; Jan. 1901-Dec. 2022. NERC EDS Centre for Environmental Data Analysis (2023).
27. Lan, X., Tans, P. & Thoning, K. NOAA Greenhouse Gas Marine Boundary Layer Reference - CO2. NOAA GML https://doi.org/10.15138/DVNP-F961 (2023).
28. Ciais, P. *et al.* Definitions and methods to estimate regional land carbon fluxes for the second phase of the REgional Carbon Cycle Assessment and Processes Project (RECCAP-2). *Geosci. Model Dev.* **15**, 1289–1316 (2022).

29. Ma, L. *et al.* Global evaluation of the Ecosystem Demography model (ED v3.0). *Geosci. Model Dev.* **15**, 1971–1994 (2022).

30. Obermeier, W. A. *et al.* Modelled land use and land cover change emissions – a spatio-temporal comparison of different approaches. *Earth Syst. Dyn.* **12**, 635–670 (2021).

31. Chini, L. *et al.* Land-use harmonization datasets for annual global carbon budgets. *Earth Syst. Sci. Data* **13**, 4175–4189 (2021).

32. Hansen, M. C. *et al.* Humid tropical forest disturbance alerts using Landsat data. *Environ. Res. Lett.* **11**, 034008 (2016).

33. Vitolo, C. *et al.* ERA5-based global meteorological wildfire danger maps. *Sci. Data* **7**, 216 (2020).

34. Besnard, S. *et al.* Memory effects of climate and vegetation affecting net ecosystem CO2 fluxes in global forests. *PLOS ONE* **14**, e0211510 (2019).

35. Ravuri, S. *et al.* Skilful precipitation nowcasting using deep generative models of radar. *Nature* **597**, 672–677 (2021).

36. Price, I. *et al.* Probabilistic weather forecasting with machine learning. *Nature* **637**, 84–90 (2025).

37. Reichstein, M. *et al.* Deep learning and process understanding for data-driven Earth system science. *Nature* **566**, 195–204 (2019).

38. Schulzweida, U. *CDO User Guide*. (2022) doi:10.5281/ZENODO.7112925.

39. Hersbach, H. *et al.* ERA5 monthly averaged data on pressure levels from 1940 to present. Copernicus Climate Change Service (C3S) Climate Data Store (CDS) https://doi.org/10.24381/CDS.6860A573 (2023).

40. He, K., Zhang, X., Ren, S. & Sun, J. Deep Residual Learning for Image Recognition. in *2016 IEEE Conference on Computer Vision and Pattern Recognition (CVPR)* 770–778 (IEEE, Las Vegas, NV, USA, 2016). doi:10.1109/CVPR.2016.90.

41. Wu, Y. & He, K. Group Normalization. in *Computer Vision – ECCV 2018* (eds Ferrari, V., Hebert, M., Sminchisescu, C. & Weiss, Y.) vol. 11217 3–19 (Springer International Publishing, Cham, 2018).

42. Hendrycks, D. & Gimpel, K. Gaussian Error Linear Units (GELUs). Preprint at https://doi.org/10.48550/arXiv.1606.08415 (2016).

43. Srivastava, N., Hinton, G., Krizhevsky, A., Sutskever, I. & Salakhutdinov, R. Dropout: A Simple Way to Prevent Neural Networks from Overfitting. *J. Mach. Learn. Res.* **15**, 1929–1958 (2014).

44. Pedregosa, F. *et al.* Scikit-learn: Machine Learning in Python. *J. Mach. Learn. Res.* **12**, 2825–2830 (2011).
45. Sohn, K. *et al.* FixMatch: Simplifying Semi-Supervised Learning with Consistency and Confidence. in *Advances in Neural Information Processing Systems* vol. 33 596–608 (Curran Associates, 2020).
46. Loshchilov, I. & Hutter, F. Decoupled Weight Decay Regularization. in *International Conference on Learning Representations* (2019).
47. Paszke, A. *et al.* PyTorch: An Imperative Style, High-Performance Deep Learning Library. in *Advances in Neural Information Processing Systems* vol. 32 8024–8035 (Curran Associates, 2019).
48. University of East Anglia Climatic Research Unit & Harris, I. C. CRU JRA v2.3: a forcings dataset of gridded land surface blend of Climatic Research Unit (CRU) and Japanese reanalysis (JRA) data; Jan. 1901-Dec. 2021. NERC EDS Centre for Environmental Data Analysis (2022).
49. Harris, I., Osborn, T. J., Jones, P. & Lister, D. Version 4 of the CRU TS monthly high-resolution gridded multivariate climate dataset. *Sci. Data* **7**, 109 (2020).

# Acknowledgements

We acknowledge the Global Carbon Project, which is responsible for the Global Carbon Budget and we thank the land modelling groups for producing and making available their model output. We thank I. Donà for discussions. P.K., P.C., S.S. and P.F. acknowledge support from the European Space Agency THRAC3E project (4000149569/25/I-AG). P.C., S.S. and P.F. also acknowledge the support from the CALIPSO (Carbon Loss in Plant Soils and Oceans) project, funded through the generosity of Schmidt Sciences, from the European Space Agency Climate Change Initiative (ESA-CCI) RECCAP2-CS project (ESA ESRIN/4000144908/24/I-LR), from ESA Carbon-RO (4000140982/23/I-EF) and from the CLARiTy project, also funded by Schmidt Sciences. P.C. received funding from the European Union's Horizon Europe research and innovation programme under grant agreement No 101081395 (EYE-CLIMA).

# Author contributions

P.K., X.G. and P.C. designed the study. X.G. and P.K. developed and trained the emulators. P.K. analysed the results and wrote the first draft of the manuscript. P.C. supervised the study and substantially revised the manuscript. X.L. contributed to the data processing. Z.L. initiated the collaboration with Microsoft Research and provided computing resources. J.B. supervised the machine-learning work at Microsoft Research. S.S. and P.F. contributed to the design of the study and provided the TRENDY data. All authors commented on the manuscript.

## Competing interests

The authors declare no competing interests.

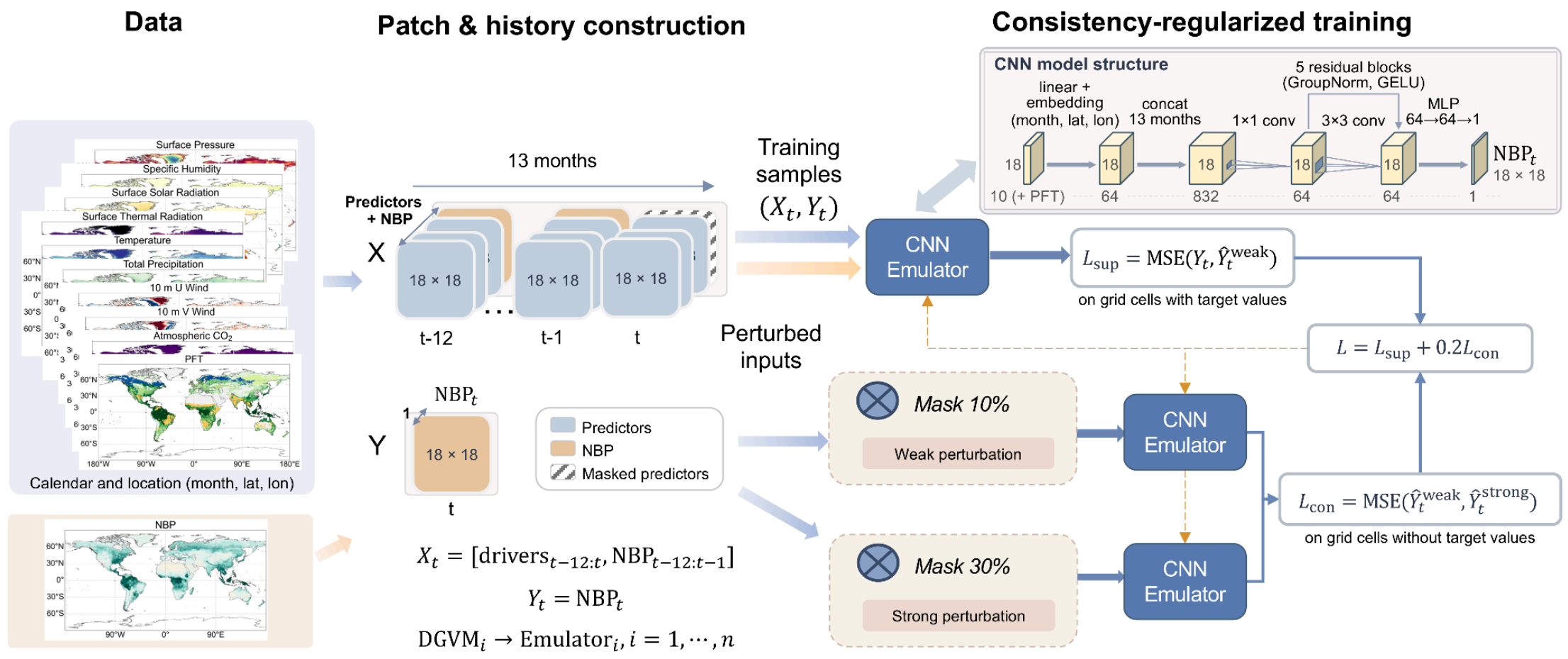

**Fig. 1 Schematic of the emulation approach.** Monthly predictor fields, atmospheric $CO_2$ and DGVM-specific plant functional type fractions (for the 12 models that provide them) are remapped to 0.5° and paired with the monthly NBP of each model. The gridded fields are divided into overlapping 18 × 18 patches and combined over 13 months. For months $t - 12$ to $t - 1$, the input contains predictors and lagged NBP. For month $t$, the input contains predictors only, with the NBP history channel masked. The convolutional network maps $X_t = [drivers_{t-12:t}, NBP_{t-12:t-1}]$, together with month, latitude and longitude, to current month $NBP_t$. A separate emulator is trained for each target model.

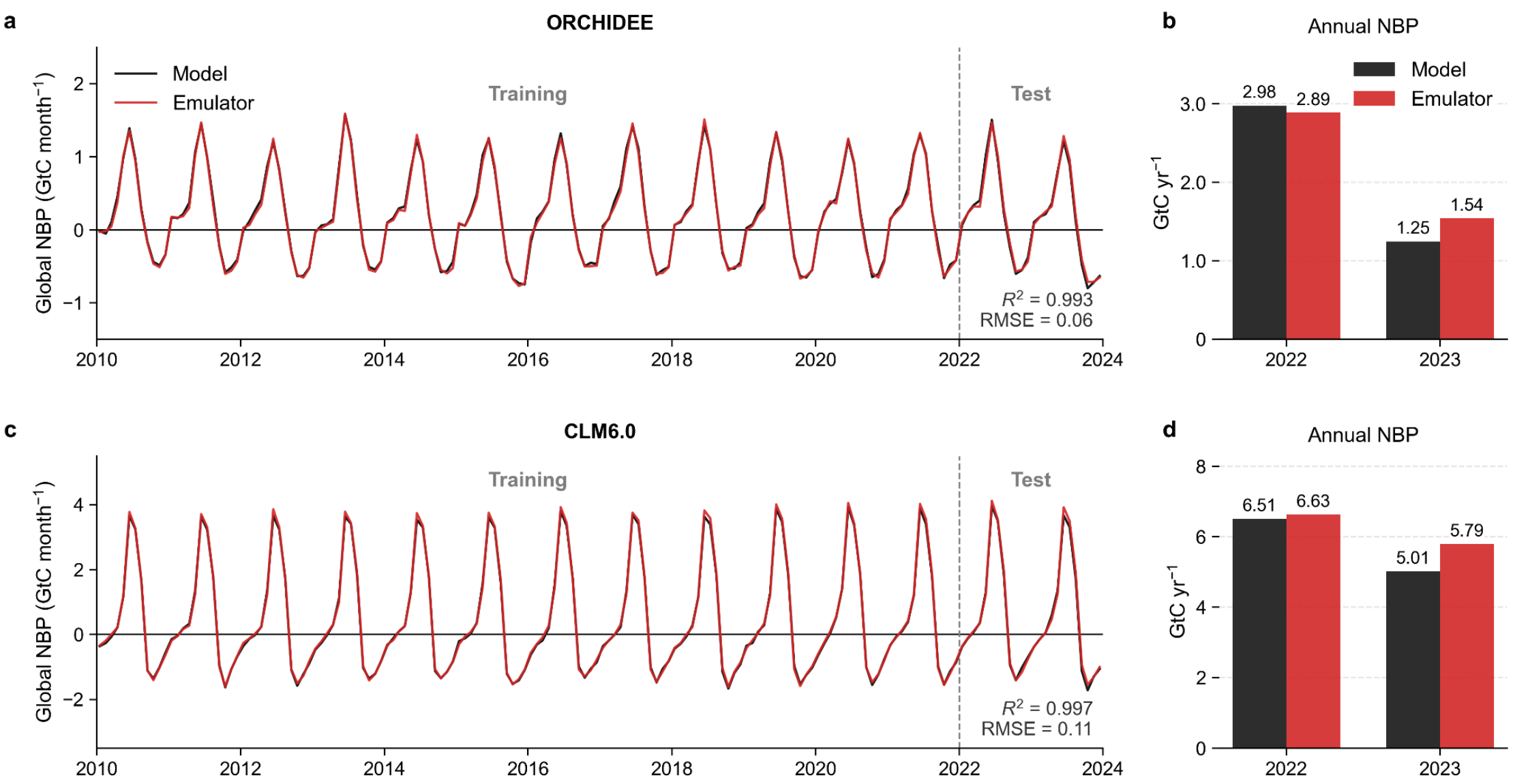

**Fig. 2 Global NBP from ORCHIDEE and CLM6.0 and from their emulators. a**,**c**, Monthly global NBP from ORCHIDEE (**a**) and CLM6.0 (**c**) (black) and from the corresponding emulators (red), 2010–2023. The dashed line marks the start of the test period (January 2022). Before 2022, the emulator output is in-sample, and from January 2022 it is predicted recursively. $R^2$ (squared Pearson correlation) and root mean square error (RMSE; GtC month$^{-1}$) are for monthly NBP in 2022–2023. **b**,**d**, Annual NBP in 2022 and 2023 from the model (black) and the emulator (red). ORCHIDEE and CLM6.0 are shown because they differ in the magnitude and seasonal amplitude of global NBP. Results for the other 17 models are shown in Supplementary Fig. 1.

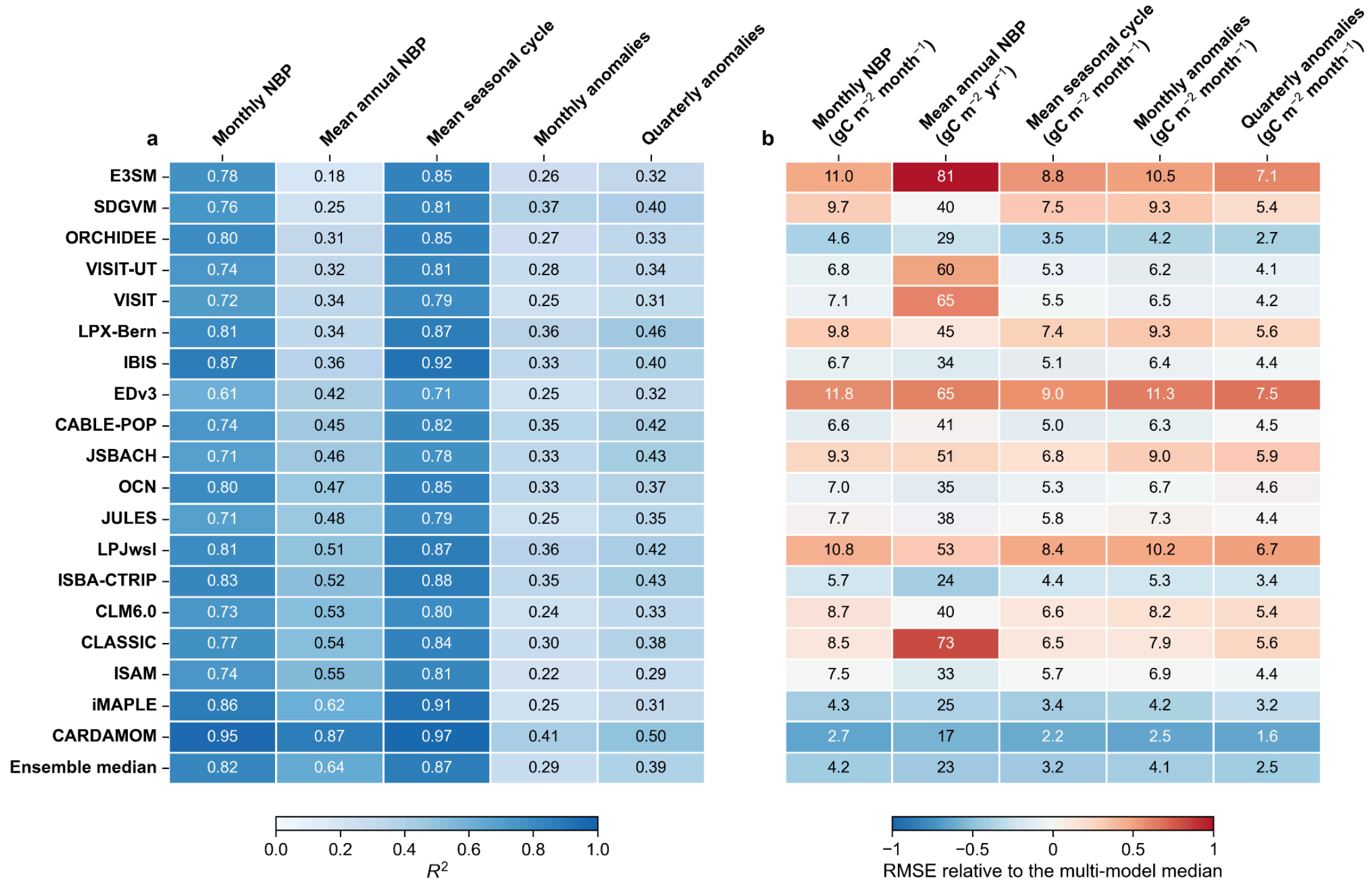

**Fig. 3 Grid-cell evaluation of the 19 emulators in the 2022–2023 test period.** a, $R^2$ between the emulator and the model. **b**, RMSE. Numbers are RMSE values; colours show (RMSE − median)/median, where the median is taken over the 19 models (blue, below the median; red, above). For monthly NBP, the mean seasonal cycle, and monthly and quarterly anomalies, $R^2$ and RMSE were calculated at each grid cell from 24 monthly, 12 calendar-month, 24 monthly and 8 quarterly values, respectively. The panels show the mean $R^2$ and the median RMSE over valid grid cells (Methods). For mean annual NBP, $R^2$ and RMSE were calculated across grid cells between the 2022–2023 mean maps of the emulator and the model. Rows are sorted by $R^2$ of mean annual NBP. Ensemble median: the median of the emulators with a value at each grid cell and month, evaluated against the median of the same models.

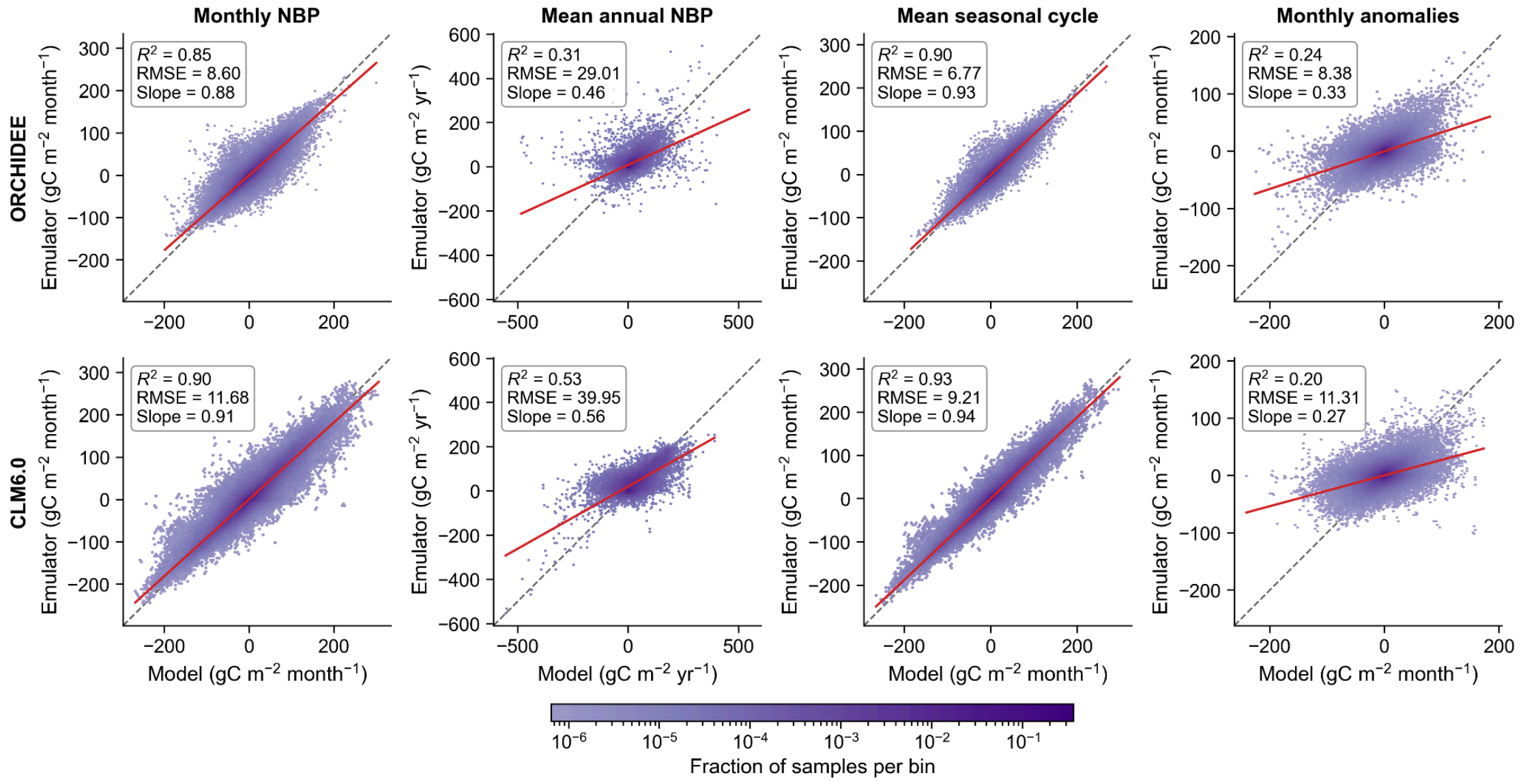

**Fig. 4 Grid-cell NBP from ORCHIDEE and CLM6.0 compared with their emulators, 2022–2023.** In each panel, all valid grid cells are pooled: 24 monthly values per grid cell for monthly NBP and monthly anomalies, 12 calendar-month values for the mean seasonal cycle, and the 2022–2023 mean for mean annual NBP. Colours show the fraction of samples in each hexagonal bin (log scale). The dashed line is the 1:1 line and the red line is the least-squares regression of the emulator on the model. $R^2$, RMSE and the regression slope are given in each panel. Results for the other 17 models are shown in Supplementary Figs. 3–5.

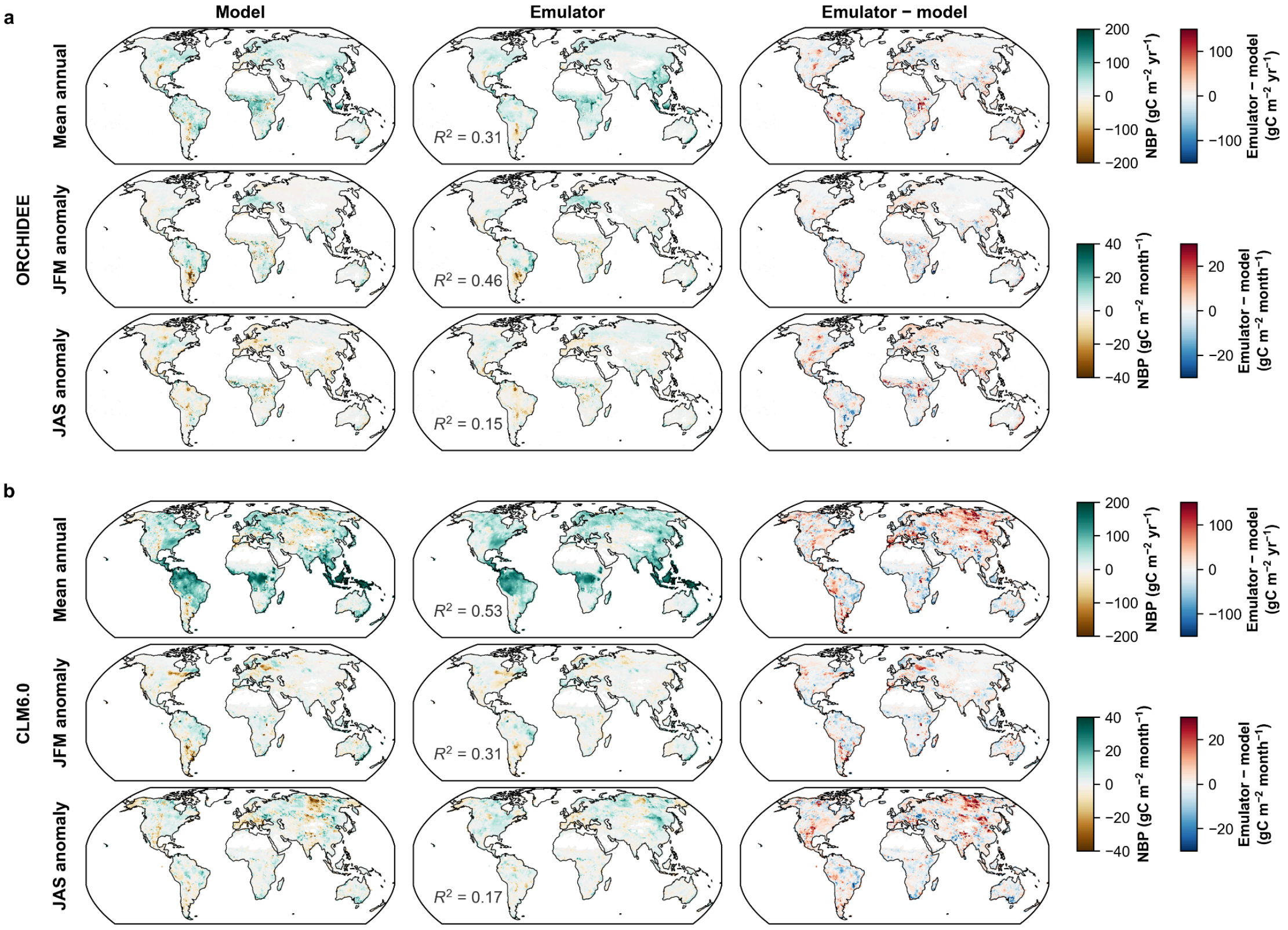

**Fig. 5 Maps of NBP from ORCHIDEE and CLM6.0 and from their emulators, 2022–2023. a**, ORCHIDEE. **b**, CLM6.0. Rows show mean annual NBP and the January–March (JFM) and July–September (JAS) anomalies, each averaged over 2022 and 2023. Columns show the model, the emulator and their difference (emulator minus model). The $R^2$ in the emulator panels is the spatial $R^2$ between the emulator and model maps over valid grid cells.

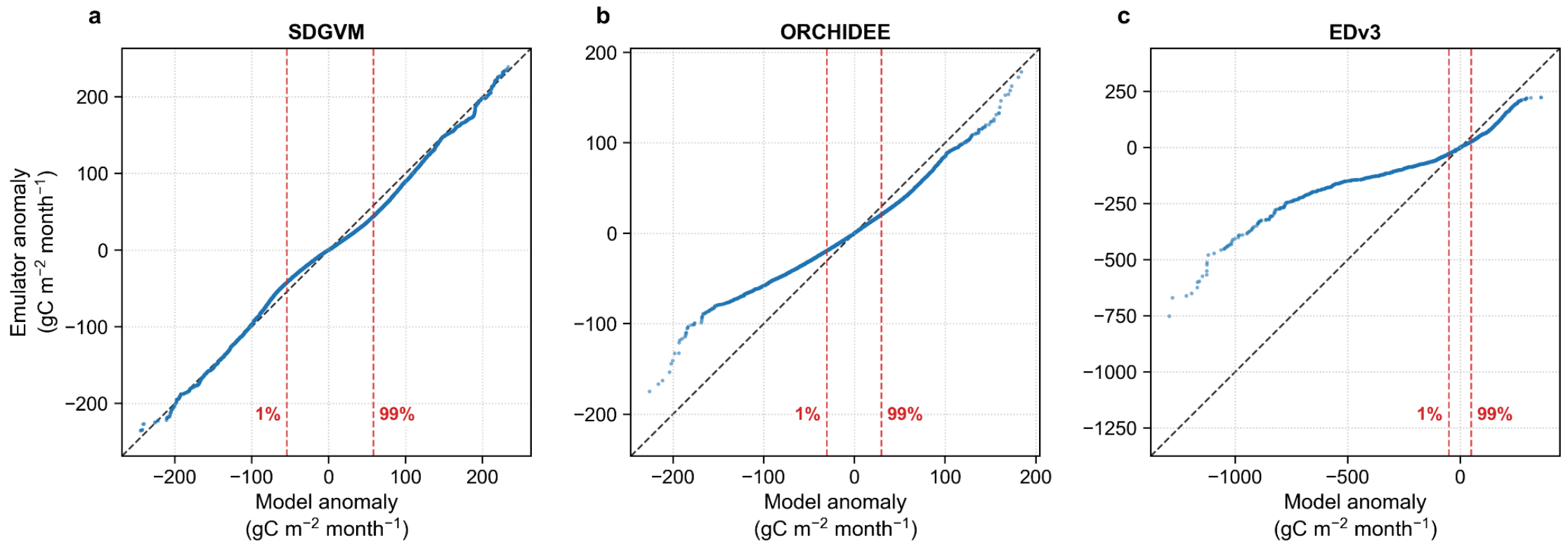

**Fig. 6 Quantile–quantile plots of monthly NBP anomalies from SDGVM, ORCHIDEE and EDv3 and from their emulators, 2022–2023.** Each point pairs the same quantile of the model anomalies (x axis) and the emulator anomalies (y axis), pooled over valid grid cells and the 24 months. The dashed line is the 1:1 line. The red vertical lines mark the 1st and 99th percentiles of the model anomalies. SDGVM, ORCHIDEE and EDv3 are shown as examples of small, intermediate and large underestimation of extreme anomalies by the emulators. Results for the other 16 models are shown in Supplementary Figs. 7–9.

# Supplement

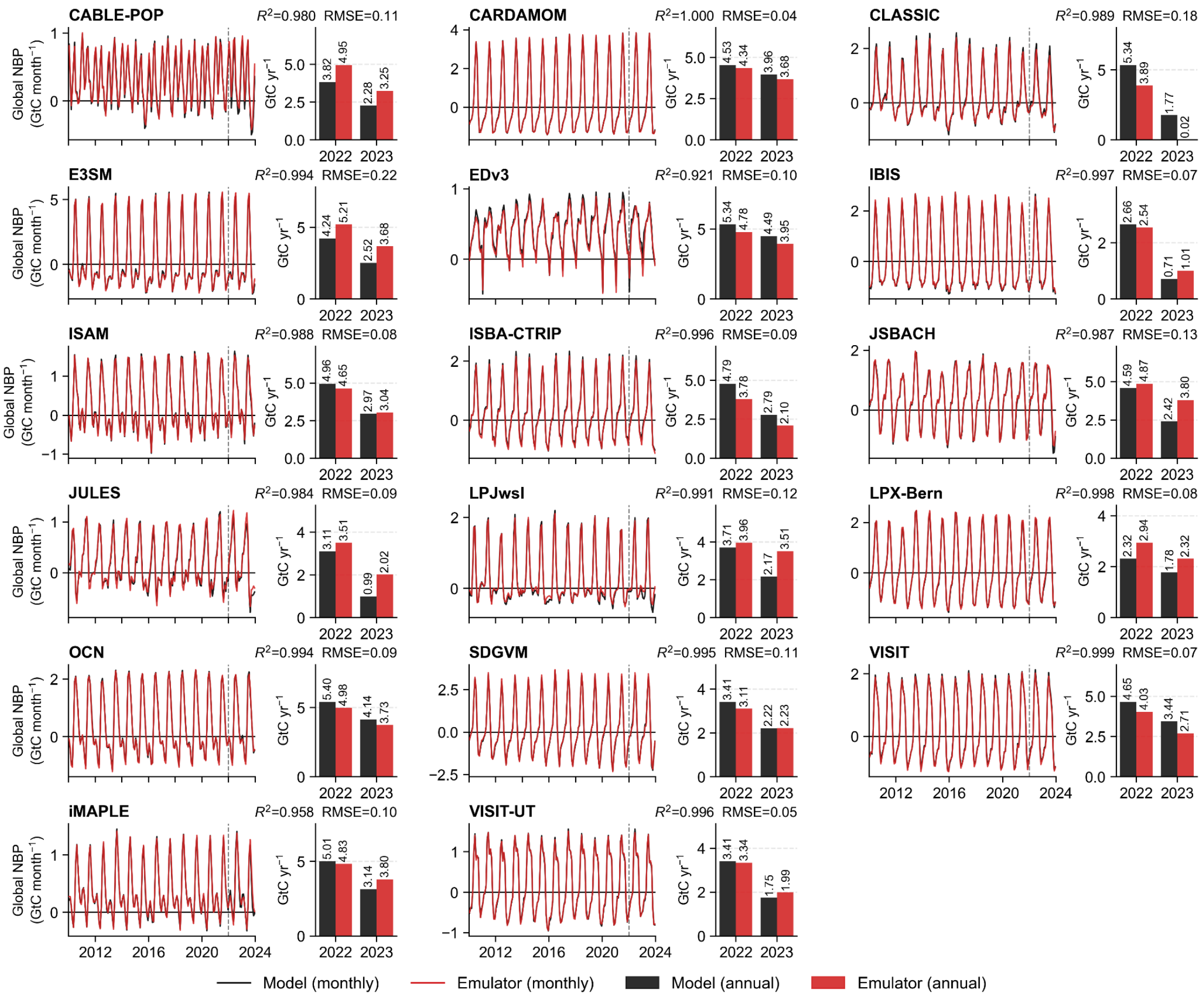

**Supplementary Fig. 1 Global NBP from the 17 models not shown in Fig. 2 and from their emulators.** For each model, the left panel shows monthly global NBP from the model (black) and the emulator (red), 2010–2023, and the right panel shows annual NBP in 2022 and 2023. The dashed line marks the start of the test period (January 2022). Before 2022, the emulator output is in-sample, and from January 2022 it is predicted recursively. $R^2$ and RMSE (GtC month$^{-1}$) are for monthly NBP in 2022–2023.

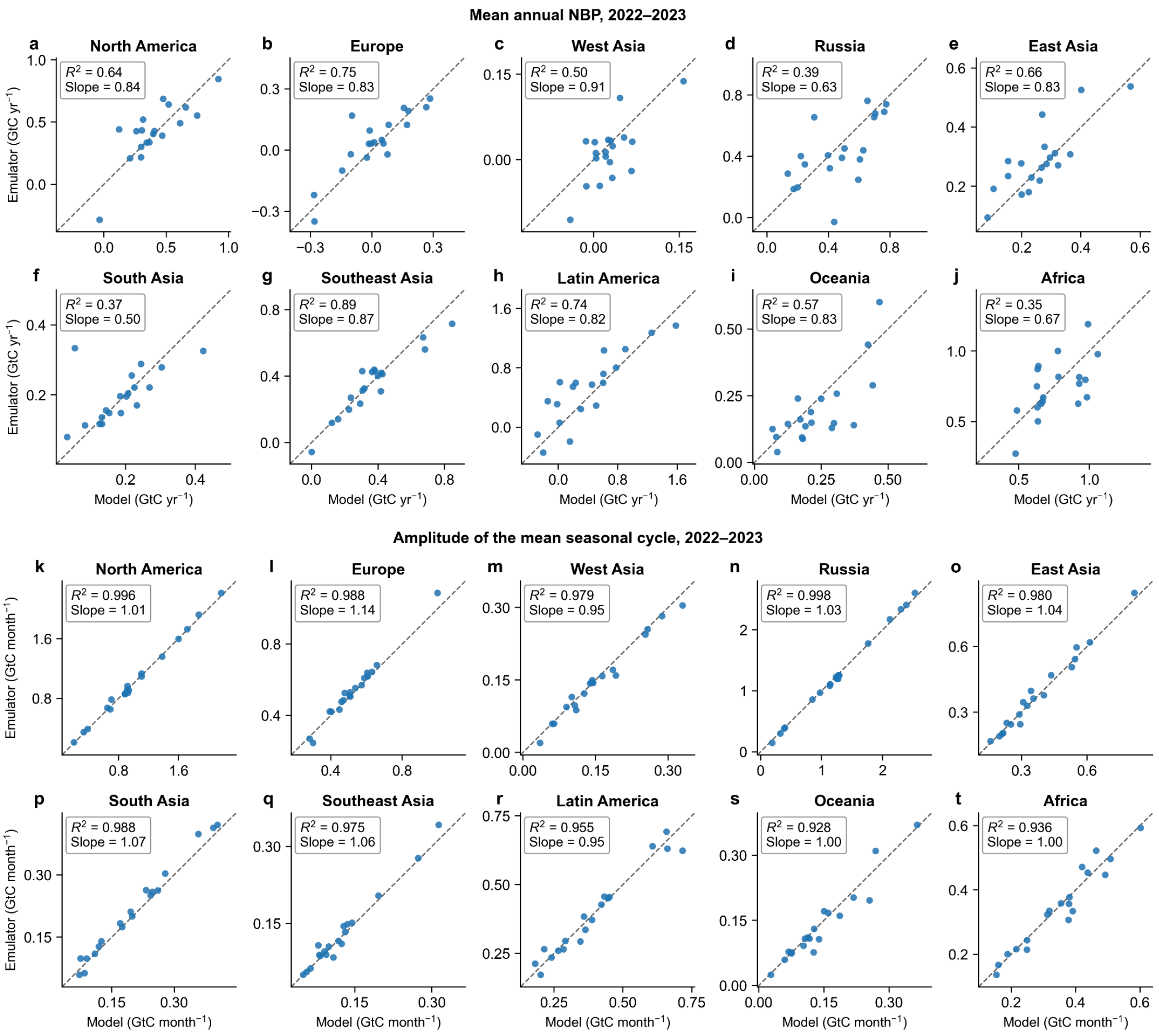

**Supplementary Fig. 2 Regional NBP from the 19 models and from their emulators, 2022–2023.** Each point is one model. **a–j**, Mean annual NBP in the ten regions. **k–t**, Amplitude of the mean seasonal cycle. The dashed line is the 1:1 line. $R^2$ and the slope of the least-squares regression of the emulator on the model are given in each panel.

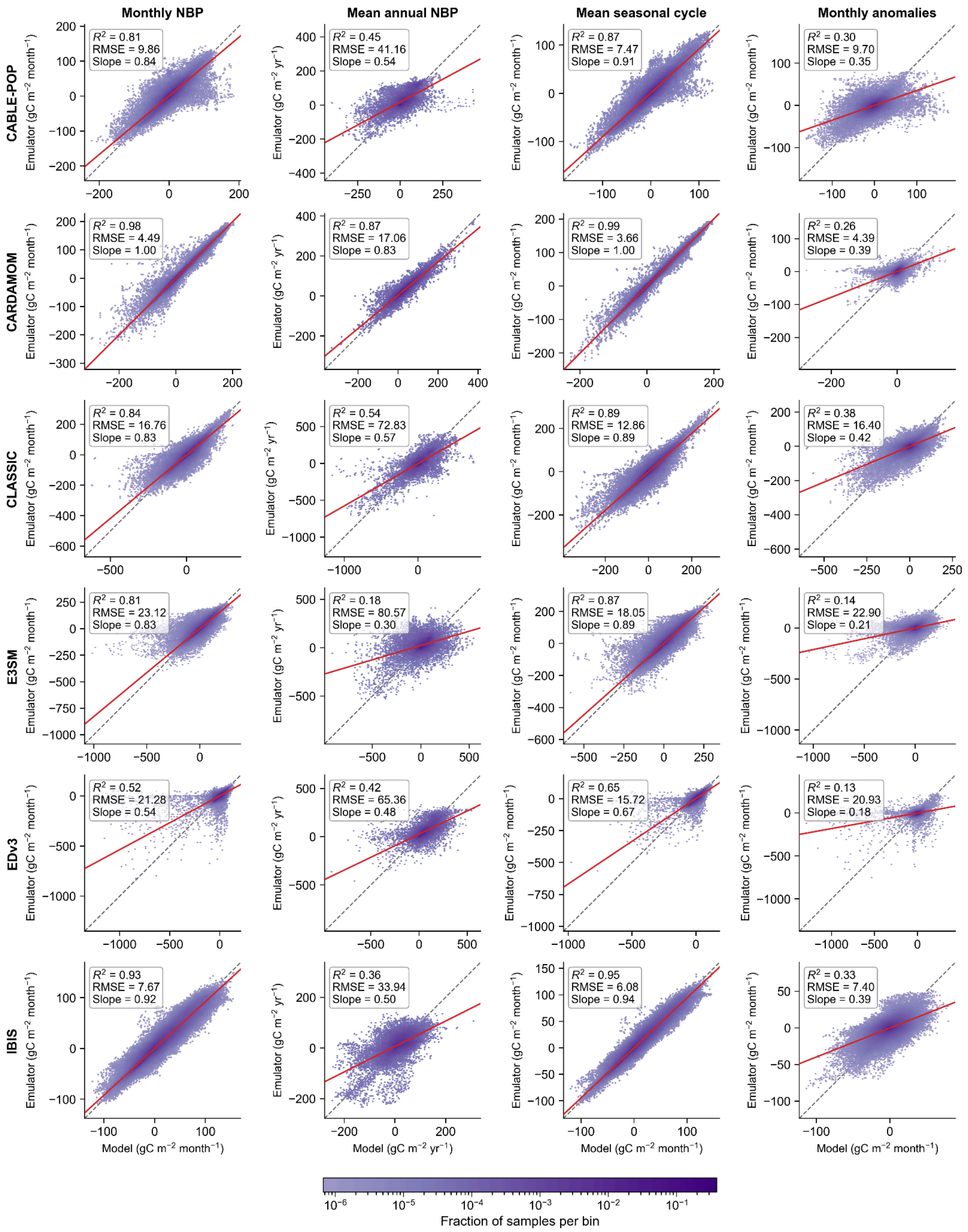

**Supplementary Fig. 3 Grid-cell NBP from CABLE-POP, CARDAMOM, CLASSIC, E3SM, EDv3 and IBIS compared with their emulators, 2022–2023.** Columns show monthly NBP, mean annual NBP, the mean seasonal cycle and monthly anomalies. In each panel, all valid grid cells are pooled: 24 monthly values per grid cell for monthly NBP and monthly anomalies, 12 calendar-month values for the mean seasonal cycle, and the 2022–2023 mean for mean annual NBP. Colours show the fraction of samples in each hexagonal bin (log scale). The dashed line is the 1:1 line and the red line is the least-squares regression of the emulator on the model. $R^2$, RMSE and the regression slope are given in each panel. The 17 models not shown in Fig. 4 are split across Supplementary Figs. 3–5.

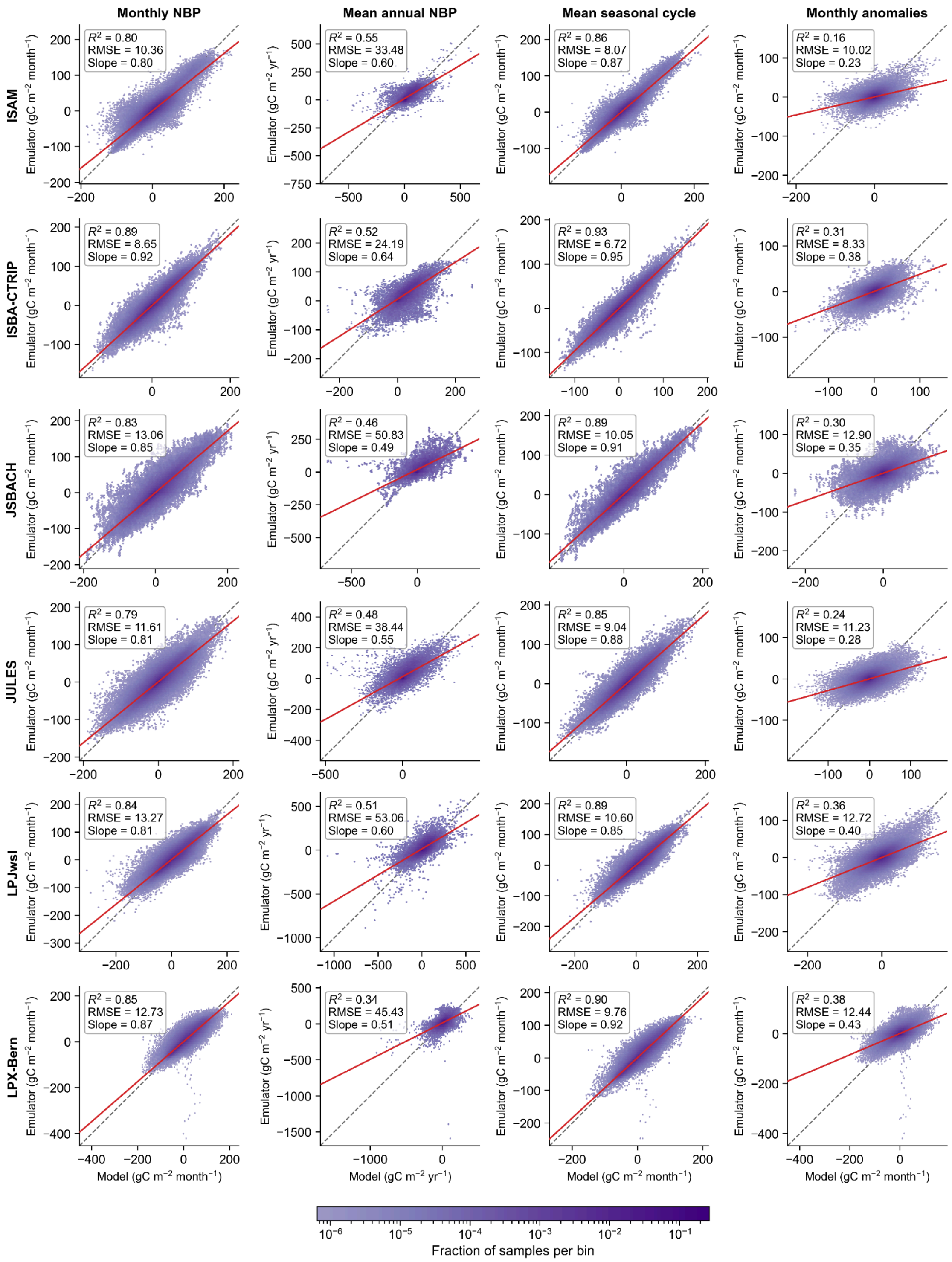

**Supplementary Fig. 4 Grid-cell NBP from ISAM, ISBA-CTRIP, JSBACH, JULES, LPJwsl and LPX-Bern compared with their emulators, 2022–2023.** Columns show monthly NBP, mean annual NBP, the mean seasonal cycle and monthly anomalies. In each panel, all valid grid cells are pooled: 24 monthly values per grid cell for monthly NBP and monthly anomalies, 12 calendar-month values for the mean seasonal cycle, and the 2022–2023 mean for mean annual NBP. Colours show the fraction of samples in each hexagonal bin (log scale). The dashed line is the 1:1 line and the red line is the least-squares regression of the emulator on the model. $R^2$, RMSE and the regression slope are given in each panel.

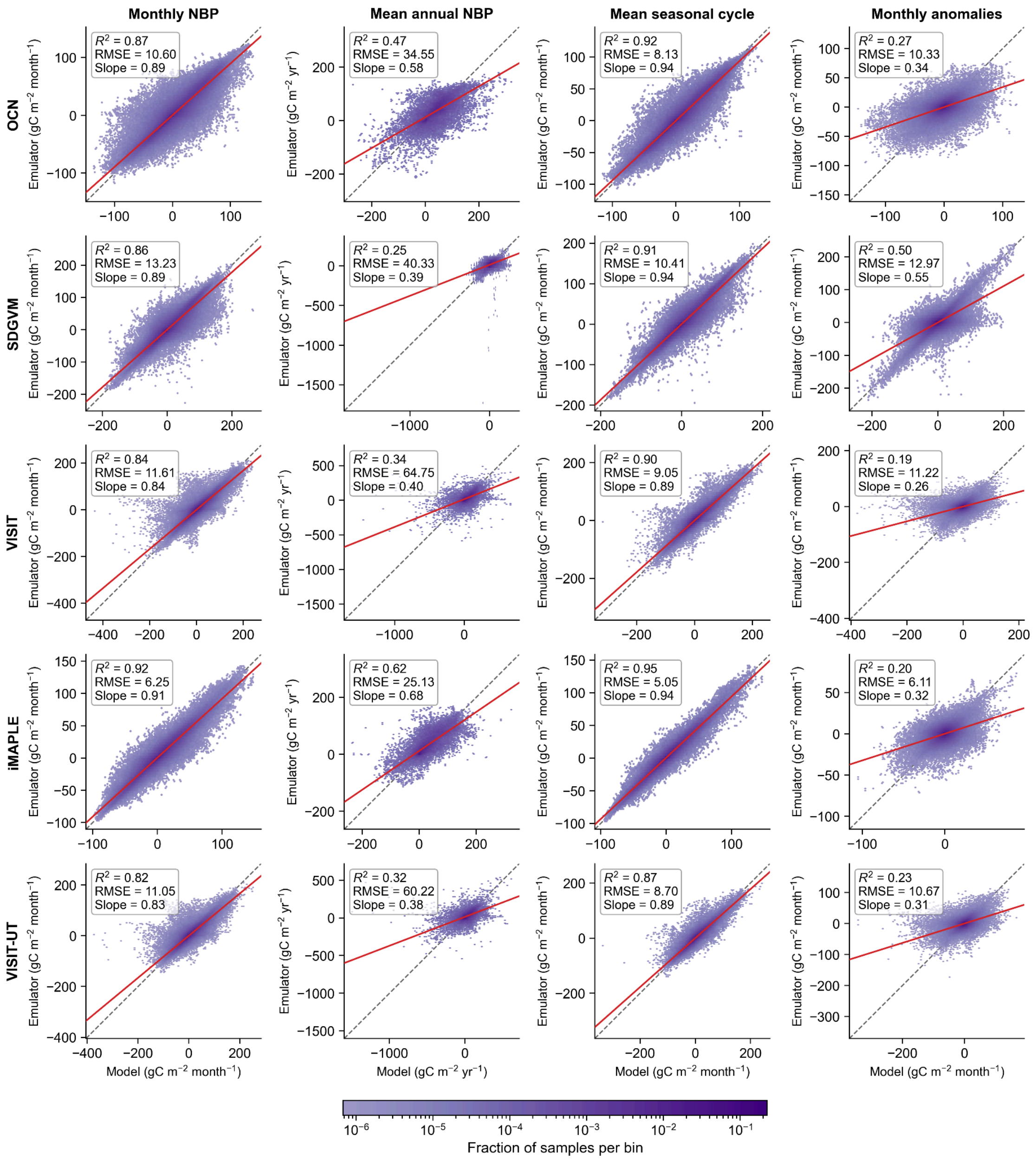

**Supplementary Fig. 5 Grid-cell NBP from OCN, SDGVM, VISIT, iMAPLE and VISIT-UT compared with their emulators, 2022–2023.** Columns show monthly NBP, mean annual NBP, the mean seasonal cycle and monthly anomalies. In each panel, all valid grid cells are pooled: 24 monthly values per grid cell for monthly NBP and monthly anomalies, 12 calendar-month values for the mean seasonal cycle, and the 2022–2023 mean for mean annual NBP. Colours show the fraction of samples in each hexagonal bin (log scale). The dashed line is the 1:1 line and the red line is the least-squares regression of the emulator on the model. $R^2$, RMSE and the regression slope are given in each panel.

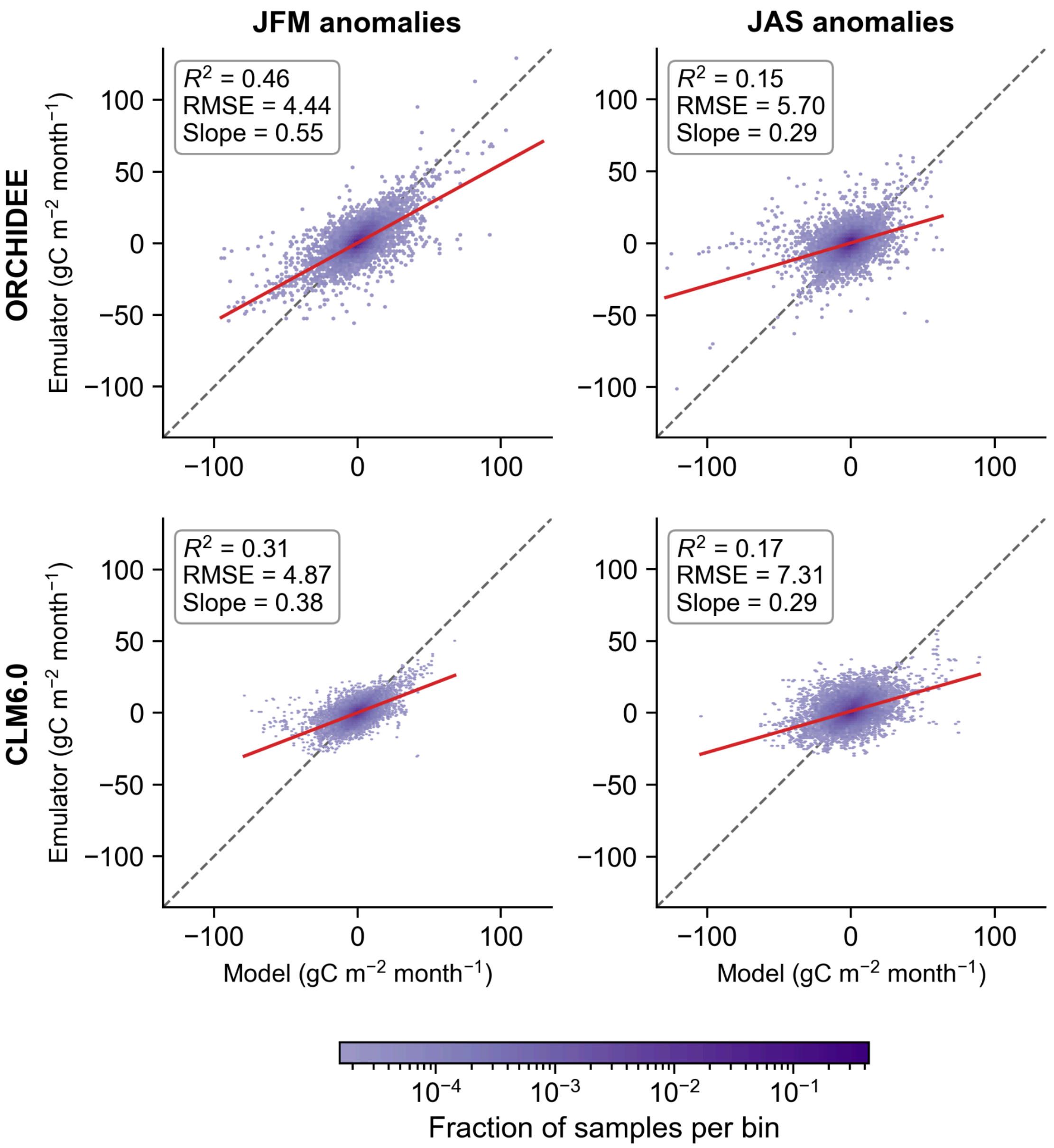

**Supplementary Fig. 6 Seasonal NBP anomalies from ORCHIDEE and CLM6.0 compared with their emulators, 2022–2023.** Each sample is one valid grid cell; samples are grouped into hexagonal bins. Values are the January–March (JFM) and July–September (JAS) anomalies averaged over 2022 and 2023. Colours show the fraction of samples in each hexagonal bin (log scale). The dashed line is the 1:1 line and the red line is the least-squares regression of the emulator on the model. $R^2$, RMSE and the regression slope are given in each panel.

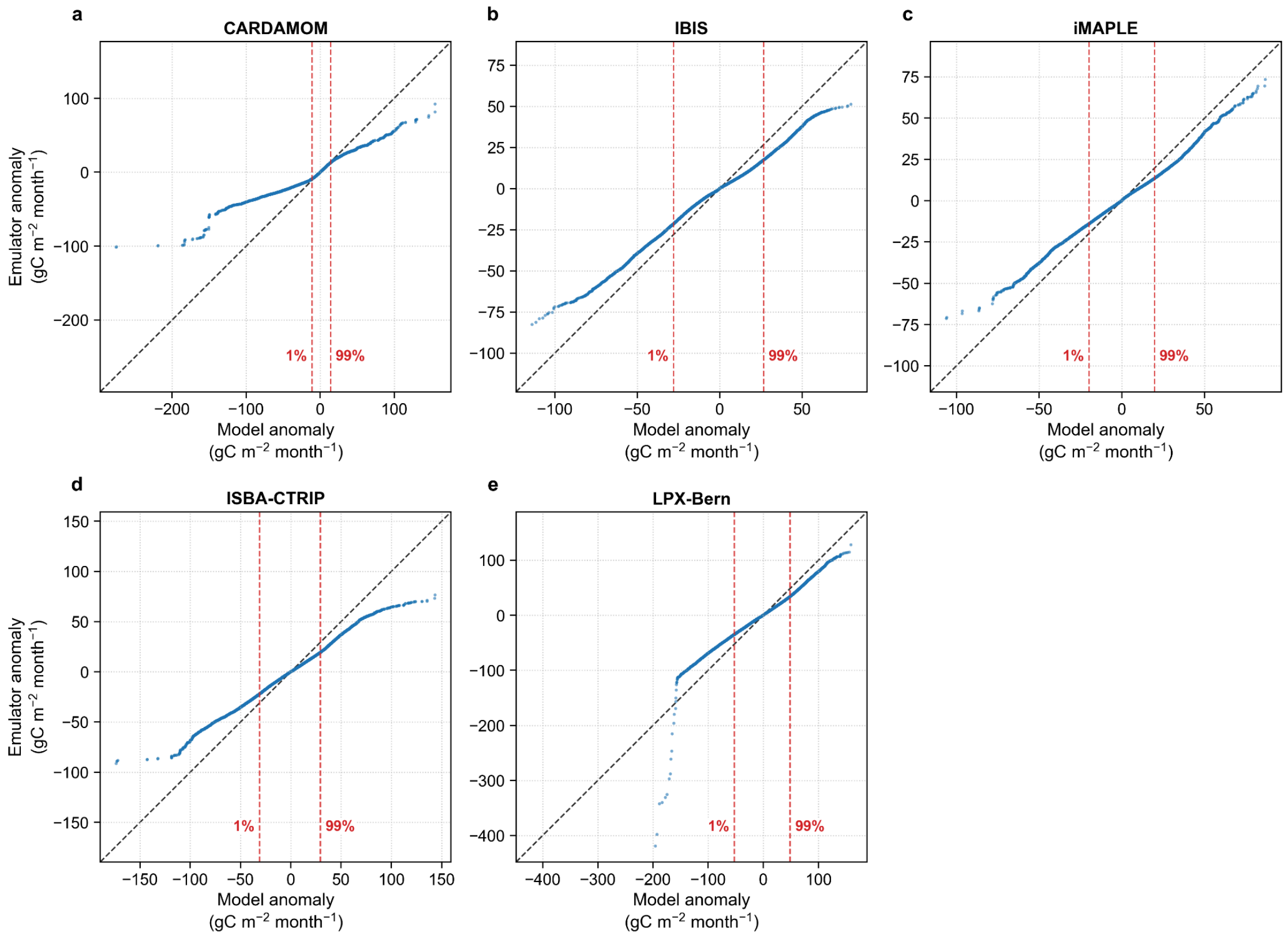

**Supplementary Fig. 7 Quantile–quantile plots of monthly NBP anomalies from CARDAMOM, IBIS, iMAPLE, ISBA-CTRIP and LPX-Bern and from their emulators, 2022–2023.** These five models have the smallest underestimation of extreme anomalies by the emulators among the 16 models not shown in Fig. 6. Each point pairs the same quantile of the model anomalies (x axis) and the emulator anomalies (y axis), pooled over valid grid cells and the 24 months. The dashed line is the 1:1 line. The red vertical lines mark the 1st and 99th percentiles of the model anomalies. The 16 models are ranked by the mean underestimation of their 1st and 99th percentiles (1 − emulator quantile/model quantile) and split into three groups (Supplementary Figs. 7–9). Panels are ordered by increasing underestimation.

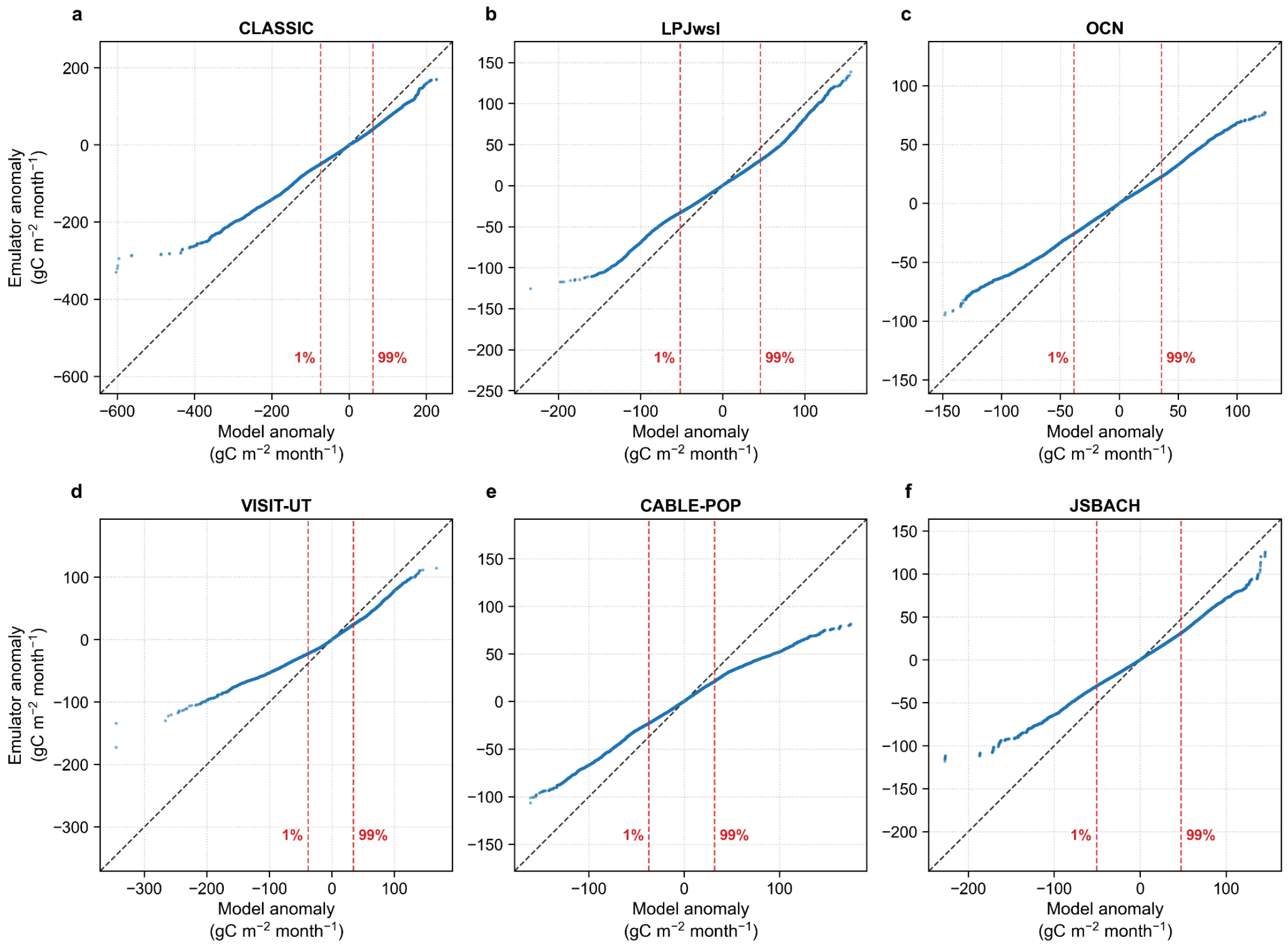

**Supplementary Fig. 8 Quantile–quantile plots of monthly NBP anomalies from CLASSIC, LPJwsl, OCN, VISIT-UT, CABLE-POP and JSBACH and from their emulators, 2022–2023.** These six models have intermediate underestimation of extreme anomalies by the emulators among the 16 models not shown in Fig. 6. Each point pairs the same quantile of the model anomalies (x axis) and the emulator anomalies (y axis), pooled over valid grid cells and the 24 months. The dashed line is the 1:1 line. The red vertical lines mark the 1st and 99th percentiles of the model anomalies. The 16 models are ranked by the mean underestimation of their 1st and 99th percentiles (1 − emulator quantile/model quantile) and split into three groups (Supplementary Figs. 7–9). Panels are ordered by increasing underestimation.

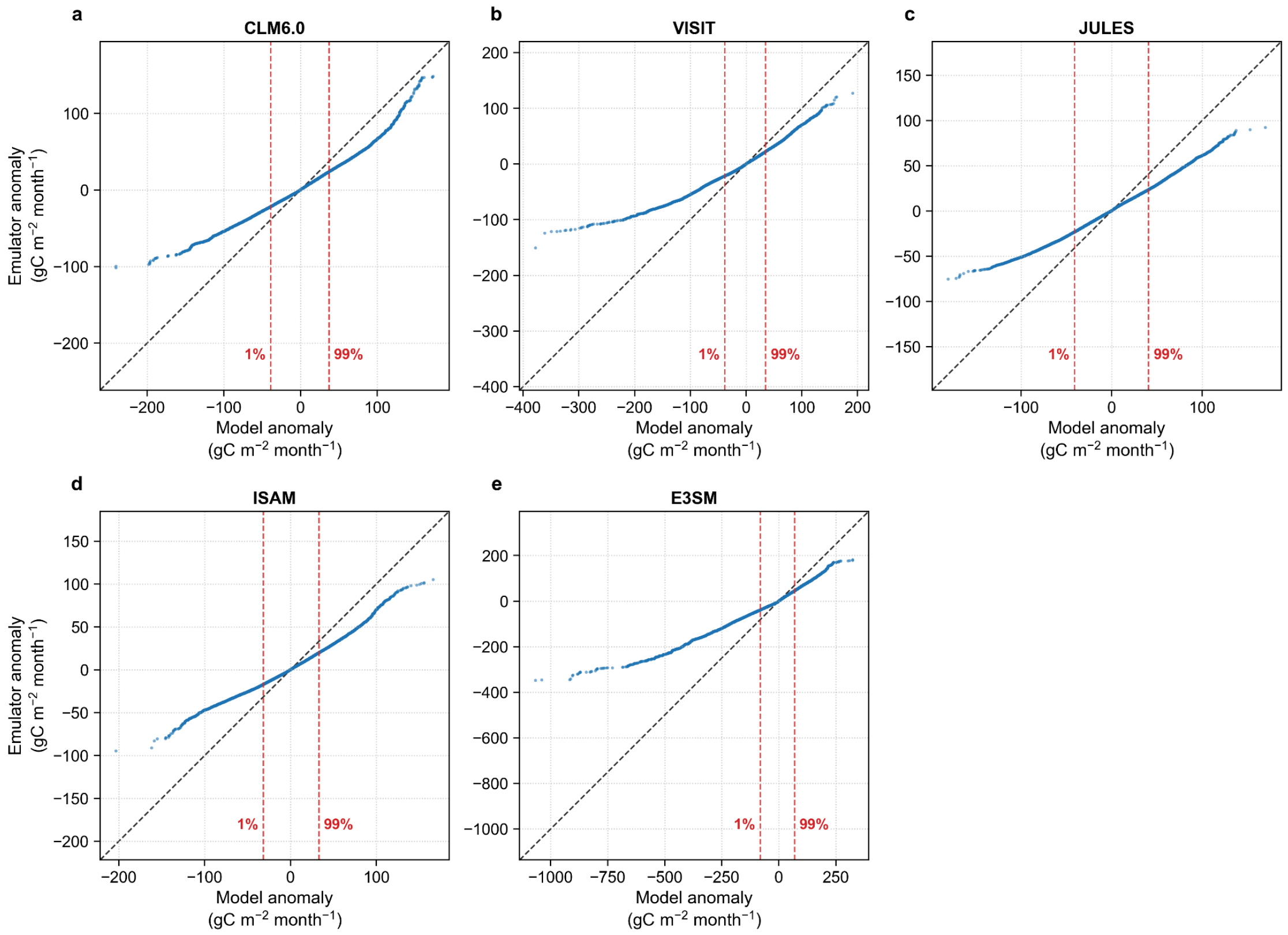

**Supplementary Fig. 9 Quantile–quantile plots of monthly NBP anomalies from CLM6.0, VISIT, JULES, ISAM and E3SM and from their emulators, 2022–2023.** These five models have the largest underestimation of extreme anomalies by the emulators among the 16 models not shown in Fig. 6. Each point pairs the same quantile of the model anomalies (x axis) and the emulator anomalies (y axis), pooled over valid grid cells and the 24 months. The dashed line is the 1:1 line. The red vertical lines mark the 1st and 99th percentiles of the model anomalies. The 16 models are ranked by the mean underestimation of their 1st and 99th percentiles (1 − emulator quantile/model quantile) and split into three groups (Supplementary Figs. 7–9). Panels are ordered by increasing underestimation.

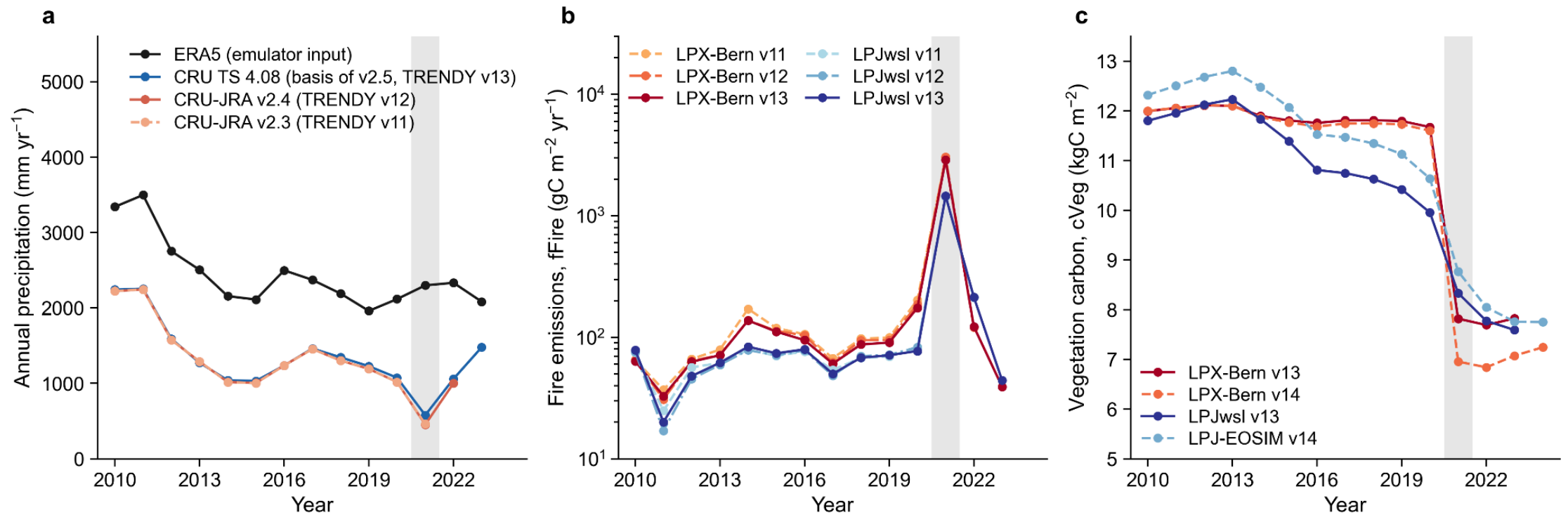

**Supplementary Fig. 10 Precipitation, fire emissions and vegetation carbon in 36 grid cells of western Venezuela, 2010–2024. a**, Annual precipitation from ERA5, CRU TS 4.08, CRU-JRA v2.4 and CRU-JRA v2.3. **b**, Fire carbon emissions simulated by LPX-Bern and LPJwsl in TRENDY v11, v12 and v13 (log scale). **c**, Vegetation carbon simulated by LPX-Bern and LPJwsl in TRENDY v13 and v14 (LPJwsl is named LPJ-EOSIM in v14). Model results are from the S2 simulation. All values are means over the 36 grid cells (6.75–9.75° N, 73.75–70.25° W), which are excluded from the evaluation (Methods). The grey band marks 2021.

**Supplementary Table 1 Models from the TRENDY v13 ensemble of the Global Carbon Budget 2024 used as emulation targets.**

| Model | Fire | Data information | References |
|---|---|---|---|
| CABLE-POP | No | Monthly NBP from each model, covering the period from Jan 2000 (Jan 2003 for CARDAMOM) to Dec 2023 | Haverd et al. (2018) |
| CLASSIC | Yes | | Melton et al. (2020), Asaadi et al. (2018) |
| CLM6.0 | Yes | | Lawrence et al. (2019) |
| EDv3 | Yes | | Moorcroft et al. (2001), Ma et al. (2022) |
| E3SM | Yes | | Yang et al. (2023), Burrows et al. (2020) |
| IBIS | Yes | | Xia et al. (2024) |
| iMAPLE | No | | Yue et al. (2024) |
| ISAM | No | | Jain et al. (2013), Meiyappan et al. (2015), Shu et al. (2020) |
| ISBA-CTRIP | Yes | | Delire et al. (2020) |
| JSBACH | Yes | | Mauritsen et al. (2019), Reick et al. (2021) |
| JULES | Yes | | Wiltshire et al. (2021), Sellar et al. (2019), Burton et al. (2019) |
| LPJwsl | Yes | | Poulter et al. (2011) |
| LPX-Bern | Yes | | Lienert and Joos (2018) |
| OCN | No | | Zaehle and Friend (2010), Zaehle et al. (2011) |
| ORCHIDEE | No | | Krinner et al. (2005), Zaehle and Friend (2010), Vuichard et al. (2019) |
| SDGVM | Yes | | Woodward and Lomas (2004), Walker et al. (2017) |
| VISIT | Yes | | Ito and Inatomi (2012), Kato et al. (2013) |
| VISIT-UT | Yes | | Ito (2019) |
| CARDAMOM | Yes[a] | | Bloom et al. (2016), Smallman et al. (2021) |

Fire indicates whether the model simulates fire, as reported in Table S1 of the Global Carbon Budget 2024 (Friedlingstein et al., 2025) or, for VISIT-UT, which is not listed there, in Table S1 of the Global Carbon Budget 2025 (Friedlingstein et al., 2026). [a]Burned area is prescribed from satellite observations, and the resulting carbon losses are simulated by the model.

**Supplementary Table 2 Input predictor variables used for emulator training.**

| Variable | Data information | Source |
|---|---|---|
| Atmospheric $CO_2$ concentration | Monthly, covering the period from Jan 2000 to Dec 2024 | NOAA: Greenhouse Gas Marine Boundary Layer Reference (Lan et al., 2023) |
| Surface pressure | Monthly, covering the period from Jan 2000 to Dec 2024 | ERA5 (Hersbach et al., 2023a, b) |
| Surface solar radiation downwards | Monthly, covering the period from Jan 2000 to Dec 2024 | ERA5 (Hersbach et al., 2023a, b) |
| Surface thermal radiation downwards | Monthly, covering the period from Jan 2000 to Dec 2024 | ERA5 (Hersbach et al., 2023a, b) |
| 2 metre temperature | Monthly, covering the period from Jan 2000 to Dec 2024 | ERA5 (Hersbach et al., 2023a, b) |
| Total precipitation | Monthly, covering the period from Jan 2000 to Dec 2024 | ERA5 (Hersbach et al., 2023a, b) |
| 10 metre U wind component | Monthly, covering the period from Jan 2000 to Dec 2024 | ERA5 (Hersbach et al., 2023a, b) |
| 10 metre V wind component | Monthly, covering the period from Jan 2000 to Dec 2024 | ERA5 (Hersbach et al., 2023a, b) |
| Specific humidity at 1000 hPa | Monthly, covering the period from Jan 2000 to Dec 2024 | ERA5 (Hersbach et al., 2023a, b) |
| NBP of the target model | Monthly, preceding 12 months; in the prediction period, NBP predicted by the emulator | TRENDY v13 S2 (Supplementary Table 1) |
| PFT fractions of each model (12 models) | Static, annual or monthly depending on the model, covering the period from 2000 to 2023 (2023 values used for 2024) | GCB 2024 (Friedlingstein et al., 2025) |
| Month | N.A. | N.A. |
| Latitude and longitude | N.A. | N.A. |

**Supplementary Table 3 Evaluation of the emulators and of climatology at the three spatial scales, 2022–2023.**

| Scale | Quantity | Emulators | Climatology |
|---|---|---|---|
| Global | Annual NBP: median absolute error (GtC $yr^{-1}$), 2022 / 2023 | 0.31 / 0.66 | 0.84 / 0.97 |
| Global | Annual NBP: number of models with an absolute error below 0.4 GtC $yr^{-1}$, 2022 / 2023 | 10 / 6 | 4 / 4 |
| Global | Monthly NBP: median $R^2$ | 0.994 | 0.979 |
| Global | Monthly NBP anomalies: median $R^2$ / median skill score | 0.81 / 0.72 | – / 0 |
| Regional | Annual NBP: median absolute error (GtC $yr^{-1}$) | 0.02–0.24 | 0.02–0.54 |
| Regional | Monthly NBP: median $R^2$ | 0.92–0.995 | 0.71–0.99 |
| Regional | Monthly NBP anomalies: median $R^2$ / median skill score | 0.32–0.75 / 0.30–0.72 | – / 0 |
| Grid cell | Map of mean annual NBP: median spatial $R^2$ | 0.46 | 0.21 |
| Grid cell | Monthly NBP: median $R^2$ | 0.77 | 0.76 |
| Grid cell | Mean seasonal cycle: median $R^2$ | 0.84 | 0.86 |
| Grid cell | Monthly NBP anomalies: median $R^2$ / median skill score | 0.30 / 0.23 | – / 0 |

Values are medians over the 19 models. For the regional scale, values are the ranges over the ten regions, and over the two years for annual NBP, of these medians. Skill scores are relative to climatology, so the skill score of climatology is zero. Climatology sets anomalies to zero, so its $R^2$ is not defined for anomalies (Methods).

**Supplementary Table 4 Bias of the emulators in global annual NBP.**

| Model | Mean bias, 2010–2021 (GtC $yr^{-1}$) | s.d. of bias, 2010–2021 (GtC $yr^{-1}$) | Bias, 2022 (GtC $yr^{-1}$) | Outside ±2 s.d., 2022 | Bias, 2023 (GtC $yr^{-1}$) | Outside ±2 s.d., 2023 | Change 2022 to 2023, model (GtC $yr^{-1}$) | Change 2022 to 2023, emulator (GtC $yr^{-1}$) |
|---|---|---|---|---|---|---|---|---|
| CABLE-POP | 0.10 | 0.16 | 1.12 | Yes | 0.97 | Yes | −1.55 | −1.70 |
| CLASSIC | −0.53 | 0.25 | −1.45 | Yes | −1.76 | Yes | −3.57 | −3.87 |
| CLM6.0 | 0.27 | 0.20 | 0.12 | No | 0.78 | Yes | −1.50 | −0.84 |
| EDv3 | −0.44 | 0.16 | −0.55 | No | −0.55 | No | −0.84 | −0.84 |
| E3SM | −0.41 | 0.35 | 0.97 | Yes | 1.15 | Yes | −1.71 | −1.53 |
| IBIS | 0.23 | 0.15 | −0.11 | Yes | 0.30 | No | −1.94 | −1.53 |
| iMAPLE | 0.08 | 0.11 | −0.18 | Yes | 0.66 | Yes | −1.87 | −1.03 |
| ISAM | −0.19 | 0.17 | −0.31 | No | 0.08 | No | −2.00 | −1.61 |
| ISBA-CTRIP | −0.60 | 0.23 | −1.01 | No | −0.69 | No | −1.99 | −1.68 |
| JSBACH | 0.09 | 0.15 | 0.29 | No | 1.39 | Yes | −2.17 | −1.07 |
| JULES | 0.06 | 0.22 | 0.40 | No | 1.02 | Yes | −2.12 | −1.49 |
| LPJwsl | 0.11 | 0.26 | 0.25 | No | 1.34 | Yes | −1.54 | −0.44 |
| LPX-Bern | 0.14 | 0.21 | 0.62 | Yes | 0.54 | No | −0.54 | −0.63 |
| OCN | −0.30 | 0.18 | −0.42 | No | −0.41 | No | −1.26 | −1.24 |
| ORCHIDEE | −0.19 | 0.09 | −0.09 | No | 0.29 | Yes | −1.73 | −1.34 |
| SDGVM | 0.23 | 0.25 | −0.31 | Yes | 0.01 | No | −1.20 | −0.88 |
| VISIT | −0.41 | 0.11 | −0.62 | No | −0.73 | Yes | −1.21 | −1.32 |
| VISIT-UT | −0.05 | 0.12 | −0.07 | No | 0.24 | Yes | −1.67 | −1.35 |
| CARDAMOM | −0.17 | 0.12 | −0.19 | No | −0.28 | No | −0.57 | −0.67 |
| Mean of 19 models | −0.10 | 0.12 | −0.08 | No | 0.23 | Yes | −1.63 | −1.32 |

Bias is emulator minus model. Outside ±2 s.d. indicates whether the bias lies outside the 2010–2021 mean bias ± 2 s.d. The last row is for the mean of the 19 models and the corresponding mean of their emulators.

**Supplementary Table 5 $R^2$, RMSE and skill scores of the emulators for global monthly NBP in 2022–2023.**

| Model | $R^2$, monthly NBP | $R^2$, monthly anomalies | RMSE, monthly anomalies 2022 (GtC month$^{-1}$) | RMSE, monthly anomalies 2023 (GtC month$^{-1}$) | SS, climatology | SS, persistence (December 2021) | SS, persistence (2021 mean) | SS, ERA5 regression | Regression $R^2$, 2010–2021 | Regression $R^2$, 2022–2023 |
|---|---|---|---|---|---|---|---|---|---|---|
| CABLE-POP | 0.980 | 0.85 | 0.101 | 0.090 | 0.62 | 0.70 | 0.80 | −0.05 | 0.39 | 0.61 |
| CLASSIC | 0.989 | 0.81 | 0.099 | 0.157 | 0.67 | 0.90 | 0.74 | 0.71 | 0.43 | 0.49 |
| CLM6.0 | 0.997 | 0.53 | 0.056 | 0.107 | 0.55 | 0.59 | 0.51 | 0.25 | 0.27 | 0.56 |
| EDv3 | 0.921 | 0.40 | 0.091 | 0.087 | 0.41 | 0.40 | 0.47 | 0.08 | 0.10 | 0.47 |
| E3SM | 0.994 | 0.43 | 0.215 | 0.195 | 0.15 | 0.26 | 0.44 | −0.26 | 0.21 | 0.34 |
| IBIS | 0.997 | 0.79 | 0.071 | 0.064 | 0.77 | 0.82 | 0.78 | 0.24 | 0.27 | 0.80 |
| iMAPLE | 0.958 | 0.35 | 0.080 | 0.103 | 0.33 | 0.34 | 0.36 | −0.03 | 0.34 | 0.61 |
| ISAM | 0.988 | 0.72 | 0.042 | 0.093 | 0.74 | 0.72 | 0.71 | 0.49 | 0.35 | 0.71 |
| ISBA-CTRIP | 0.996 | 0.83 | 0.069 | 0.040 | 0.80 | 0.88 | 0.88 | 0.68 | 0.35 | 0.63 |
| JSBACH | 0.987 | 0.80 | 0.060 | 0.164 | 0.67 | 0.81 | 0.70 | 0.36 | 0.21 | 0.65 |
| JULES | 0.984 | 0.84 | 0.058 | 0.114 | 0.73 | 0.74 | 0.74 | 0.31 | 0.33 | 0.69 |
| LPJwsl | 0.991 | 0.75 | 0.043 | 0.134 | 0.56 | 0.61 | 0.54 | 0.41 | 0.26 | 0.47 |
| LPX-Bern | 0.998 | 0.82 | 0.052 | 0.065 | 0.72 | 0.72 | 0.72 | 0.91 | 0.33 | 0.02 |
| OCN | 0.994 | 0.69 | 0.041 | 0.099 | 0.67 | 0.78 | 0.67 | 0.47 | 0.23 | 0.53 |
| ORCHIDEE | 0.993 | 0.82 | 0.050 | 0.056 | 0.77 | 0.85 | 0.85 | 0.53 | 0.25 | 0.49 |
| SDGVM | 0.995 | 0.84 | 0.101 | 0.107 | 0.76 | 0.80 | 0.77 | 0.75 | 0.14 | 0.26 |
| VISIT | 0.999 | 0.88 | 0.028 | 0.045 | 0.83 | 0.90 | 0.83 | 0.90 | 0.45 | 0.40 |
| VISIT-UT | 0.996 | 0.93 | 0.028 | 0.034 | 0.92 | 0.96 | 0.92 | 0.93 | 0.51 | 0.47 |
| CARDAMOM | 1.000 | 0.89 | 0.027 | 0.034 | 0.94 | 0.91 | 0.87 | 0.92 | 0.18 | 0.02 |
| Mean of 19 models | 0.999 | 0.97 | 0.021 | 0.047 | 0.89 | 0.91 | 0.90 | 0.72 | 0.40 | 0.84 |

SS, skill score relative to the baseline given in the column heading (Methods). Regression $R^2$ refers to the ERA5 regression baseline. The last row is for the mean of the 19 models and the corresponding mean of their emulators.

**Supplementary Table 6 Emulator predictions for 2024 compared with TRENDY v14.**

| Model | Grid cells | $R^2$ between v14 and v13 monthly anomalies, 2010–2021 | Annual anomaly 2024, v14 (GtC $yr^{-1}$) | Annual anomaly 2024, emulator (GtC $yr^{-1}$) | Same sign | $R^2$, monthly anomalies 2024 | RMSE, monthly anomalies 2024 (GtC $month^{-1}$) | SS, climatology | SS, persistence (2023 mean, v13) | SS, ERA5 regression |
|---|---|---|---|---|---|---|---|---|---|---|
| CABLE-POP | 53,138 | 0.919 | −0.22 | −0.69 | Yes | 0.75 | 0.059 | 0.56 | 0.83 | 0.81 |
| CLASSIC | 61,867 | 0.832 | −2.62 | −2.43 | Yes | 0.31 | 0.179 | 0.66 | 0.30 | 0.57 |
| CLM6.0 | 62,614 | 0.741 | −0.56 | 0.32 | No | 0.71 | 0.107 | 0.23 | 0.13 | 0.30 |
| EDv3 | 29,020 | 0.934 | −0.61 | −0.90 | Yes | 0.12 | 0.100 | −0.01 | −0.36 | −0.16 |
| E3SM | 57,788 | 0.879 | −1.33 | −0.37 | Yes | 0.13 | 0.160 | −0.16 | −1.58 | −0.51 |
| IBIS | 53,783 | 0.972 | −1.53 | −1.53 | Yes | 0.19 | 0.086 | 0.67 | 0.03 | −0.42 |
| iMAPLE | 57,985 | 0.974 | −0.10 | 0.01 | No | 0.38 | 0.042 | 0.33 | 0.77 | 0.64 |
| ISAM | 49,750 | 0.947 | −0.53 | 0.18 | No | 0.33 | 0.101 | −0.24 | −0.60 | −0.05 |
| JSBACH | 57,020 | 0.889 | −0.45 | −0.03 | Yes | 0.02 | 0.146 | −1.03 | −0.07 | −0.17 |
| JULES | 51,448 | 0.511 | −1.71 | −1.28 | Yes | 0.34 | 0.095 | 0.72 | 0.30 | 0.50 |
| LPJwsl | 53,475 | 0.899 | −0.54 | −1.44 | Yes | 0.47 | 0.089 | −0.35 | −0.13 | −0.43 |
| LPX-Bern | 55,678 | 0.993 | −0.26 | −0.23 | Yes | 0.57 | 0.077 | 0.58 | 0.57 | 0.46 |
| ORCHIDEE | 57,461 | 0.946 | −0.57 | −0.37 | Yes | 0.29 | 0.086 | 0.40 | 0.43 | 0.46 |
| SDGVM | 60,644 | 0.703 | −0.81 | 0.74 | No | 0.64 | 0.172 | 0.24 | 0.21 | 0.35 |
| VISIT | 51,050 | 0.984 | −0.31 | −0.09 | Yes | 0.66 | 0.049 | 0.53 | 0.46 | 0.83 |
| VISIT-UT | 47,040 | 0.859 | −0.78 | −0.49 | Yes | 0.42 | 0.059 | 0.56 | 0.08 | 0.57 |
| CARDAMOM | 53,187 | 0.661 | 1.45 | 0.83 | Yes | 0.25 | 0.092 | 0.54 | −0.37 | 0.50 |
| Mean of 17 models | | 0.957 | −0.68 | −0.46 | Yes | 0.31 | 0.042 | 0.66 | 0.21 | 0.59 |
| Mean of 10 models | | 0.968 | −0.51 | −0.51 | Yes | 0.31 | 0.034 | 0.64 | 0.55 | 0.72 |

Grid cells, number of valid grid cells (Methods) that also have complete data in TRENDY v14 and in the emulator output, with the variability threshold applied to TRENDY v14. OCN and ISBA-CTRIP are not in TRENDY v14. LPJwsl is named LPJ-EOSIM in TRENDY v14 (Methods). The last two rows are for the means of the 17 models and of the 10 models whose monthly anomalies were highly correlated between versions ($R^2$ above 0.88: CABLE-POP, EDv3, IBIS, iMAPLE, ISAM, JSBACH, LPJwsl, LPX-Bern, ORCHIDEE and VISIT), and the corresponding means of their emulators. With $R^2$ thresholds of 0.85, 0.90 and 0.95 instead of 0.88 (12, 8 and 4 models), the skill score of the subset mean relative to persistence was 0.43, 0.38 and 0.25.

## References

Asaadi, A., Arora, V. K., Melton, J. R., & Bartlett, P. (2018). An improved parameterization of leaf area index (LAI) seasonality in the Canadian Land Surface Scheme (CLASS) and Canadian Terrestrial Ecosystem Model (CTEM) modelling framework. *Biogeosciences*, *15*(22), 6885-6907. https://doi.org/10.5194/bg-15-6885-2018

Bloom, A. A., Exbrayat, J.-F., van der Velde, I. R., Feng, L., & Williams, M. (2016). The decadal state of the terrestrial carbon cycle: Global retrievals of terrestrial carbon allocation, pools, and residence times. *Proceedings of the National Academy of Sciences*, *113*(5), 1285-1290. https://doi.org/10.1073/pnas.1515160113

Burrows, S. M., Maltrud, M., Yang, X., Zhu, Q., Jeffery, N., Shi, X., Ricciuto, D., Wang, S., Bisht, G., Tang, J., Wolfe, J., Harrop, B. E., Singh, B., Brent, L., Baldwin, S., Zhou, T., Cameron-Smith, P., Keen, N., Collier, N.,…Leung, L. R. (2020). The DOE E3SM v1.1 Biogeochemistry Configuration: Description and Simulated Ecosystem-Climate Responses to Historical Changes in Forcing. *Journal of Advances in Modeling Earth Systems*, *12*(9), e2019MS001766. https://doi.org/10.1029/2019MS001766

Burton, C., Betts, R., Cardoso, M., Feldpausch, T. R., Harper, A., Jones, C. D., Kelley, D. I., Robertson, E., & Wiltshire, A. (2019). Representation of fire, land-use change and vegetation dynamics in the Joint UK Land Environment Simulator vn4.9 (JULES). *Geoscientific Model Development*, *12*(1), 179-193. https://doi.org/10.5194/gmd-12-179-2019

Delire, C., Séférian, R., Decharme, B., Alkama, R., Calvet, J.-C., Carrer, D., Gibelin, A.-L., Joetzjer, E., Morel, X., Rocher, M., & Tzanos, D. (2020). The Global Land Carbon Cycle Simulated With ISBA-CTRIP: Improvements Over the Last Decade. *Journal of Advances in Modeling Earth Systems*, *12*(9), e2019MS001886. https://doi.org/10.1029/2019MS001886

Friedlingstein, P., O'Sullivan, M., Jones, M. W., Andrew, R. M., Hauck, J., Landschützer, P., Le Quéré, C., Li, H., Luijkx, I. T., Olsen, A., Peters, G. P., Peters, W., Pongratz, J., Schwingshackl, C., Sitch, S., Canadell, J. G., Ciais, P., Jackson, R. B., Alin, S. R.,…Zeng, J. (2025). Global Carbon Budget 2024. *Earth System Science Data*, *17*(3), 965-1039. https://doi.org/10.5194/essd-17-965-2025

Friedlingstein, P., O'Sullivan, M., Jones, M. W., Andrew, R. M., Bakker, D. C. E., Hauck, J., Landschützer, P., Le Quéré, C., Li, H., Luijkx, I. T., Peters, G. P., Peters, W., Pongratz, J., Schwingshackl, C., Sitch, S., Canadell, J. G., Ciais, P., Aas, K., Alin, S. R.,…Zeng, J. (2026). Global Carbon Budget 2025. *Earth System Science Data*, *18*(5), 3211-3288. https://doi.org/10.5194/essd-18-3211-2026

Haverd, V., Smith, B., Nieradzik, L., Briggs, P. R., Woodgate, W., Trudinger, C. M., Canadell, J. G., & Cuntz, M. (2018). A new version of the CABLE land surface model (Subversion revision r4601) incorporating land use and land cover change, woody vegetation demography, and a novel optimisation-based approach to plant coordination of photosynthesis. *Geoscientific Model Development*, *11*(7), 2995-3026. https://doi.org/10.5194/gmd-11-2995-2018

Hersbach, H., Bell, B., Berrisford, P., Biavati, G., Horányi, A., Muñoz Sabater, J., Nicolas, J., Peubey, C., Radu, R., Rozum, I., Schepers, D., Simmons, A., Soci, C., Dee, D., & Thépaut, J.-N. (2023a). *ERA5 monthly averaged data on single levels from 1940 to present*. Copernicus Climate Change Service (C3S) Climate Data Store (CDS). https://doi.org/10.24381/cds.f17050d7

Hersbach, H., Bell, B., Berrisford, P., Biavati, G., Horányi, A., Muñoz Sabater, J., Nicolas, J., Peubey, C., Radu, R., Rozum, I., Schepers, D., Simmons, A., Soci, C., Dee, D., & Thépaut, J.-N. (2023b). *ERA5 monthly averaged data on pressure levels from 1940 to present*. Copernicus Climate Change Service (C3S) Climate Data Store (CDS). https://doi.org/10.24381/cds.6860a573

Ito, A. (2019). Disequilibrium of terrestrial ecosystem CO2 budget caused by disturbance-induced emissions and non-CO2 carbon export flows: a global model assessment. *Earth System Dynamics*, *10*(4), 685-709. https://doi.org/10.5194/esd-10-685-2019

Ito, A., & Inatomi, M. (2012). Use of a process-based model for assessing the methane budgets of global terrestrial ecosystems and evaluation of uncertainty. *Biogeosciences*, *9*(2), 759-773. https://doi.org/10.5194/bg-9-759-2012

Jain, A. K., Meiyappan, P., Song, Y., & House, J. I. (2013). CO2 emissions from land-use change affected more by nitrogen cycle, than by the choice of land-cover data. *Global Change Biology*, *19*(9), 2893-2906. https://doi.org/10.1111/gcb.12207

Kato, E., Kinoshita, T., Ito, A., Kawamiya, M., & Yamagata, Y. (2013). Evaluation of spatially explicit emission scenario of land-use change and biomass burning using a process-based biogeochemical model. *Journal of Land Use Science*, *8*(1), 104-122. https://doi.org/10.1080/1747423X.2011.628705

Krinner, G., Viovy, N., de Noblet-Ducoudré, N., Ogée, J., Polcher, J., Friedlingstein, P., Ciais, P., Sitch, S., & Prentice, I. C. (2005). A dynamic global vegetation model for studies of the coupled atmosphere-biosphere system. *Global Biogeochemical Cycles*, *19*(1), GB1015. https://doi.org/10.1029/2003GB002199

Lan, X., Tans, P., Thoning, K., & NOAA Global Monitoring Laboratory. (2023). *NOAA Greenhouse Gas Marine Boundary Layer Reference - CO2*. NOAA GML. https://doi.org/10.15138/DVNP-F961

Lawrence, D. M., Fisher, R. A., Koven, C. D., Oleson, K. W., Swenson, S. C., Bonan, G., Collier, N., Ghimire, B., van Kampenhout, L., Kennedy, D., Kluzek, E., Lawrence, P. J., Li, F., Li, H., Lombardozzi, D., Riley, W. J., Sacks, W. J., Shi, M., Vertenstein, M.,…Zeng, X. (2019). The Community Land Model Version 5: Description of New Features, Benchmarking, and Impact of Forcing Uncertainty. *Journal of Advances in Modeling Earth Systems*, *11*(12), 4245-4287. https://doi.org/10.1029/2018MS001583

Lienert, S., & Joos, F. (2018). A Bayesian ensemble data assimilation to constrain model parameters and land-use carbon emissions. *Biogeosciences*, *15*(9), 2909-2930. https://doi.org/10.5194/bg-15-2909-2018

Ma, L., Hurtt, G., Ott, L., Sahajpal, R., Fisk, J., Lamb, R., Tang, H., Flanagan, S., Chini, L., Chatterjee, A., & Sullivan, J. (2022). Global evaluation of the Ecosystem Demography model (ED v3.0). *Geoscientific Model Development*, *15*(5), 1971-1994. https://doi.org/10.5194/gmd-15-1971-2022

Mauritsen, T., Bader, J., Becker, T., Behrens, J., Bittner, M., Brokopf, R., Brovkin, V., Claussen, M., Crueger, T., Esch, M., Fast, I., Fiedler, S., Fläschner, D., Gayler, V., Giorgetta, M., Goll, D. S., Haak, H., Hagemann, S., Hedemann, C.,…Roeckner, E. (2019). Developments in the MPI-M Earth System Model version 1.2 (MPI-ESM1.2) and Its Response to Increasing CO2. *Journal of Advances in Modeling Earth Systems*, *11*(4), 998-1038. https://doi.org/10.1029/2018MS001400

Meiyappan, P., Jain, A. K., & House, J. I. (2015). Increased influence of nitrogen limitation on CO2 emissions from future land use and land use change. *Global Biogeochemical Cycles*, *29*(9), 1524-1548. https://doi.org/10.1002/2015GB005086

Melton, J. R., Arora, V. K., Wisernig-Cojoc, E., Seiler, C., Fortier, M., Chan, E., & Teckentrup, L. (2020). CLASSIC v1.0: the open-source community successor to the Canadian Land Surface Scheme (CLASS) and the Canadian Terrestrial Ecosystem Model (CTEM) – Part 1: Model framework and site-level performance. *Geoscientific Model Development*, *13*(6), 2825-2850. https://doi.org/10.5194/gmd-13-2825-2020

Moorcroft, P. R., Hurtt, G. C., & Pacala, S. W. (2001). A method for scaling vegetation dynamics: the ecosystem demography model (ED). *Ecological Monographs*, *71*(4), 557-586. https://doi.org/10.1890/0012-9615(2001)071[0557:AMFSVD]2.0.CO;2

Poulter, B., Frank, D. C., Hodson, E. L., & Zimmermann, N. E. (2011). Impacts of land cover and climate data selection on understanding terrestrial carbon dynamics and the CO2 airborne fraction. *Biogeosciences*, *8*(8), 2027-2036. https://doi.org/10.5194/bg-8-2027-2011

Reick, C. H., Gayler, V., Goll, D., Hagemann, S., Heidkamp, M., & Nabel, J. E. M. S. (2021). JSBACH 3 - The land component of the MPI Earth System Model: documentation of version 3.2. *Berichte zur Erdsystemforschung*, *240*. https://doi.org/10.17617/2.3279802

Sellar, A. A., Jones, C. G., Mulcahy, J. P., Tang, Y., Yool, A., Wiltshire, A., O'Connor, F. M., Stringer, M., Hill, R., Palmieri, J., Woodward, S., de Mora, L., Kuhlbrodt, T., Rumbold, S. T., Kelley, D. I., Ellis, R., Johnson, C. E., Walton, J., Abraham, N. L.,…Zerroukat, M. (2019). UKESM1: Description and Evaluation of the U.K. Earth System Model. *Journal of Advances in Modeling Earth Systems*, *11*(12), 4513-4558. https://doi.org/10.1029/2019MS001739

Shu, S., Jain, A. K., Koven, C. D., & Mishra, U. (2020). Estimation of Permafrost SOC Stock and Turnover Time Using a Land Surface Model With Vertical Heterogeneity of Permafrost Soils. *Global Biogeochemical Cycles*, *34*(11), e2020GB006585. https://doi.org/10.1029/2020GB006585

Smallman, T. L., Milodowski, D. T., Neto, E. S., Koren, G., Ometto, J., & Williams, M. (2021). Parameter uncertainty dominates C-cycle forecast errors over most of Brazil for the 21st century. *Earth System Dynamics*, *12*(4), 1191-1237. https://doi.org/10.5194/esd-12-1191-2021

Vuichard, N., Messina, P., Luyssaert, S., Guenet, B., Zaehle, S., Ghattas, J., Bastrikov, V., & Peylin, P. (2019). Accounting for carbon and nitrogen interactions in the global terrestrial ecosystem model ORCHIDEE (trunk version, rev 4999): multi-scale evaluation of gross primary

production. *Geoscientific Model Development*, *12*(11), 4751-4779. https://doi.org/10.5194/gmd-12-4751-2019

Walker, A. P., Quaife, T., van Bodegom, P. M., De Kauwe, M. G., Keenan, T. F., Joiner, J., Lomas, M. R., MacBean, N., Xu, C., Yang, X., & Woodward, F. I. (2017). The impact of alternative trait-scaling hypotheses for the maximum photosynthetic carboxylation rate (Vcmax) on global gross primary production. *New Phytologist*, *215*(4), 1370-1386. https://doi.org/10.1111/nph.14623

Wiltshire, A. J., Burke, E. J., Chadburn, S. E., Jones, C. D., Cox, P. M., Davies-Barnard, T., Friedlingstein, P., Harper, A. B., Liddicoat, S., Sitch, S., & Zaehle, S. (2021). JULES-CN: a coupled terrestrial carbon–nitrogen scheme (JULES vn5.1). *Geoscientific Model Development*, *14*(4), 2161-2186. https://doi.org/10.5194/gmd-14-2161-2021

Woodward, F. I., & Lomas, M. R. (2004). Vegetation dynamics – simulating responses to climatic change. *Biological Reviews*, *79*(3), 643-670. https://doi.org/10.1017/S1464793103006419

Xia, X., Ren, P., Wang, X., Liu, D., Chen, X., Dan, L., He, B., He, H., Ju, W., Liang, M., Lu, X., Peng, J., Qin, Z., Xia, J., Zheng, B., Wei, J., Yue, X., Yu, G., Piao, S., & Yuan, W. (2024). The carbon budget of China: 1980–2021. *Science Bulletin*, *69*(1), 114-124. https://doi.org/10.1016/j.scib.2023.11.016

Yang, X., Thornton, P., Ricciuto, D., Wang, Y., & Hoffman, F. (2023). Global evaluation of terrestrial biogeochemistry in the Energy Exascale Earth System Model (E3SM) and the role of the phosphorus cycle in the historical terrestrial carbon balance. *Biogeosciences*, *20*(14), 2813-2836. https://doi.org/10.5194/bg-20-2813-2023

Yue, X., Zhou, H., Tian, C., Ma, Y., Hu, Y., Gong, C., Zheng, H., & Liao, H. (2024). Development and evaluation of the interactive Model for Air Pollution and Land Ecosystems (iMAPLE) version 1.0. *Geoscientific Model Development*, *17*(11), 4621-4642. https://doi.org/10.5194/gmd-17-4621-2024

Zaehle, S., Ciais, P., Friend, A. D., & Prieur, V. (2011). Carbon benefits of anthropogenic reactive nitrogen offset by nitrous oxide emissions. *Nature Geoscience*, *4*(9), 601-605. https://doi.org/10.1038/ngeo1207

Zaehle, S., & Friend, A. D. (2010). Carbon and nitrogen cycle dynamics in the O-CN land surface model: 1. Model description, site-scale evaluation, and sensitivity to parameter estimates. *Global Biogeochemical Cycles*, *24*(1), GB1005. https://doi.org/10.1029/2009GB003521